\documentclass[english,a4paper,11pt]{article}
\usepackage[margin=3cm]{geometry}
\usepackage[utf8]{inputenc}

\usepackage{babel}
\usepackage{bm}
\usepackage[T1]{fontenc}
\usepackage{amsmath,amsfonts,amssymb,amsthm}
\usepackage{booktabs}
\usepackage[final]{graphicx}
\usepackage{float}
\usepackage[round]{natbib}
\usepackage{color}

\newcommand{\Fisher}{\mathcal I}
\newcommand{\bL}{\mathbf{L}}

\title{Portmanteau Goodness-of-Fit Tests in the Presence of Simple, Mixed, and Multiple Seasonal Autocorrelation}

\author{
	{Matteo \ Grigoletto}$^{*,\dag}$,
	\  {Francesco\ Lisi}$^{*,\ddag}$\\
	$^{*}$\small{\emph{Department  of Statistical Sciences,  University of Padua, Italy}} \\
\small{$\dag$ \ {\it matteo.grigoletto@unipd.it},  $\ddag$ \ {\it francesco.lisi@unipd.it} } }

\begin{document}
	\maketitle

%
%
\begin{abstract}

\noindent Standard diagnostic procedures for assessing the goodness of fit of linear models include tests of the null hypothesis of no residual autocorrelation against the alternative of linear dependence.
The literature proposes several portmanteau tests for residual autocorrelation in SARMA models, mainly focusing on two cases: short-term and single-seasonal autocorrelation.
Since many time series exhibit multiple seasonal patterns, whose periodic components may interact with one another, diagnostic tests for residual autocorrelation in a multi-seasonal framework are needed. However, the literature still lacks portmanteau tests specifically designed for multiple seasonal autocorrelation.
This paper addresses this gap by extending classical portmanteau tests to settings with multiple seasonalities. In addition, the proposed approach jointly tests for \textcolor{black}{both} short-term and multiple seasonal residual autocorrelation.\\
Test statistics and their asymptotic distributions are defined. Then, Monte Carlo simulations are used to evaluate the performance of the tests and the consistency with the expected results is assessed using statistical tests.
The results suggest that the proposed extension can serve as a useful goodness-of-fit diagnostic tool for the class of mSARIMA models.
\textcolor{black}{An application to road traffic data is also provided.}

\hspace{1cm}
\end{abstract}

\section{Introduction}
\label{sec:1}

Portmanteau tests in time series analysis are used to assess whether a series, or the residuals of a fitted model, still exhibits significant autocorrelation. The term "portmanteau" reflects the fact that these tests combine information from multiple lags of the autocorrelation function into a single overall statistic.\\
In a linear context, the main interest lies in adequately describing and fully accounting for the serial dependence structure of the data. This requires consideration of the autocorrelation structure of the time series, which can be efficiently represented by a \textcolor{black}{stationary, invertible and non-redundant}  ARMA$(p,q)$ process of the form:
$$\phi(B)\, Y_t=\theta(B)\, \varepsilon_t,$$ 
where $\varepsilon_t \sim \mathcal{N}(0,\sigma^2)$ \textcolor{black}{is a WN} and
$\phi(B)=(1-\phi_1 B -\ldots -\phi_p B^p)$ and $\theta(B)=(1-\theta_1 B- \ldots- \theta_q B^q)$ are the usual autoregressive and moving average polynomials, with $B$ being the lag operator.\\
\textcolor{black}{Let $\bm\beta=(\phi_1,\dots,\phi_p,\ \theta_1,\dots,\theta_q)$ be the true parameter values.  
Then the model's estimated residuals are given by
$\hat \varepsilon_t=y_t-\sum_{i=1}^p \hat \phi_i y_{t-i}+\sum_{j=1}^q
\hat \theta_j \hat \varepsilon_{t-j}$ while the residual
autocorrelation at lag $l$ is
}
\begin{equation}
  \hat \rho_l=\sum_{t=1}^{n-l} \hat \varepsilon_{t}\hat
  \varepsilon_{t-l}/ \sum_{t=1}^{n} \hat \varepsilon_{t}^2
  \hspace{0.7cm}(l=1,2,\ldots)\ .
  \label{eq:rhohat}
\end{equation}
Diagnostic checking of standard models requires verifying that the residuals are at least uncorrelated.
When this condition holds, the fitted ARMA$(p,q)$ model has been correctly identified and all the linear dependence in the time series has been fully captured. Conversely, the presence of residual autocorrelation suggests that potentially useful information in the data has been overlooked, which may compromise the reliability of estimates and forecasts.\\
Testing for residual autocorrelation is therefore a crucial step in the model estimation process. Building on the seminal work of \cite{boxpierce}, who introduced a portmanteau test for the hypothesis that the autocorrelation coefficients at the first $m$ lags of the residual series are all zero, numerous portmanteau tests have been developed  \citep{ljungbox, Monti_1994}. 
Many variants of the original test have subsequently been proposed, extending the framework to account for seasonality \citep{McLeod_1978, Mahdi_2016}, to detect nonlinearity in time series \citep{castle_hendry_2010}, and to incorporate GARCH effects \citep{mahdi_fisher_2024}, long memory, antipersistence \citep{chen_deo_2004}, and independence \citep{shao_2009, baragona_etal_2022, baragona_etal_2024}.\\
Multivariate extensions of the original test, suitable as diagnostic tools for vector autoregressive moving average (VARMA) models, have also been proposed and studied \citep{Hosking_1980, Poskitt_Tremayne, Francq_Raisi_2007, Mainassara_Amir_2020, DeGoojer_2023}.\\
In this work, we focus on univariate portmanteau tests and propose two further generalizations, each addressing a specific issue not yet considered in the literature:
\begin{enumerate}
\item the first generalization enables a joint test of the hypothesis of no short-term or seasonal residual autocorrelation for a seasonal model. This is relevant because these two hypotheses are typically tested separately, leading to inference \textcolor{black}{procedures} for which the overall significance level is no longer well defined;
\item the second generalization addresses the case of multiple seasonal autocorrelation. As high-frequency intraday data become increasingly common, multiple seasonality patterns are also more frequently encountered. This calls for new modeling approaches \citep{Delivera_etal_2011, Svetunkov_2023, Lisi_Grigoletto, Bandara_etal_2025} as well as new diagnostic tools. Indeed, applying a traditional test separately for each periodicity present in the data leads, once again, to an undefined significance level.
\end{enumerate}
To address these points, we propose extensions of existing test statistics that consider $m$ generic lags of the autocorrelation function of the residuals that is, lags that are not necessarily consecutive or equally spaced. 
\textcolor{black}{We provide the generalized test statistics as well as a proof of their asymptotic distributions. Finally, we evaluate their performance through simulation studies. In addition, unlike most previous studies, the significance of the results concerning the empirical size of the tests and their asymptotic distribution is assessed using formal statistical tests.}\\
The remainder of the paper is organized as follows: Section 2 reviews the existing literature on univariate portmanteau tests; Section 3 introduces the generalized portmanteau test and illustrates its application in the case of multiple seasonality and mixed (i.e., non-seasonal and seasonal) residual autocorrelation. Section 4 reports a Monte Carlo study examining the size, power, and asymptotic distribution of the test as well as the dependence on $m$ and $n$; \textcolor{black}{Section 5 shows an application to road traffic data.} Section 6 concludes.

\section{Literature review}

\subsection{Portmanteau tests for short-term residual autocorrelation}

Here and throughout the remainder of the paper, we use a more general
notation than that originally adopted in the literature, allowing us
to treat the generalizations that form the core of this article
without repeatedly redefining the various test statistics.

Let us therefore consider a $k$-dimensional vector of lags $\bL = (l_1,\ldots,l_k)$ .
Univariate portmanteau tests of goodness-of-fit analyze the autocorrelation of the residual time series of an estimated $ARMA(p,q)$ model through the hypothesis system
\begin{eqnarray}
	\begin{cases}
		H_0: & \rho_{l_1}=\rho_{l_2}=\ldots=\rho_{l_k}=0 \\
		H_1: & \overline{H}_0
	\end{cases}\qquad\ .
	\label{sys_hyp}
\end{eqnarray}
The null hypothesis implies no residual autocorrelation \textcolor{black}{at the considered lags} and, thus, a correctly specified model.\\
The original portmanteau tests were introduced with $k=m$ and
$l_1=1,\ldots,l_m=m$ , therefore testing for no short-term residual
autocorrelation up to lag $m$.\\
To test the hypothesis system (\ref{sys_hyp}), in a seminal work, \cite{boxpierce} proposed the test statistic
\begin{equation}
	Q_{BP}(\bL)=n \sum_{i=1}^k \hat \rho_{l_i}^2\qquad ,
	\label{BP_test}
\end{equation}
where $\hat \rho_{l_i}$ is the estimated $l_i$-order autocorrelation coefficient for the model's residual time series.
They argued that, under the null hypothesis, $Q_{BP}(\bL)$ follows,
asymptotically, a $\chi^2_\nu$ distribution with $\nu = k - \eta$,
where $\eta = p+q$ is the number of estimated parameters.\\
Later, \cite{ljungbox} showed that, for sample sizes commonly found in practice, the significance level of the test improves if $Q_{BP}(\bL)$ is replaced by the asymptotically equivalent statistic
\begin{equation}
	Q_{LB}(\bL)=n\,(n+2)\, \sum_{i=1}^k \frac{\hat \rho_{l_i}^2}{n-l_i}\qquad .
	\label{LB_test}
\end{equation}
\cite{ljung} studied the behavior of the Ljung-Box test statistic for different choices of the maximum number of lags $k=m$ and found that the properties of the statistic $Q_{LB}(\bL)$ can be improved by choosing a small value for $m$.\\
\cite{Monti_1994} proposed a variant of the original test, based on the sum of squared partial autocorrelations of the residuals, $\hat\pi_{l_i}$:
\begin{equation}
	Q_{M}(\bL)=n\, (n+2)\, \sum_{i=1}^k \frac{\hat \pi_{l_i}^2}{n-l_i} \ .
	\label{M_test}
\end{equation}
The asymptotic distribution of $ Q_{M}$ is, again, $\chi^2_\nu$ with $\nu = k - \eta$ and $\eta = p+q$. The main advantage of $Q_{M}(\bL)$ is that it is more powerful when the estimated model underestimates the moving average order.\\
\cite{pena_rodriguez} observed that testing the adequacy of an ARMA$(p,q)$ model is equivalent to testing the hypothesis that the Toeplitz matrix of the  (standardized) sample autocorrelations
\begin{equation}
	\hat{\bf{R}}(\bL)=
	\begin{pmatrix}
		1 & \hat \rho_{l_1} & \cdots &\hat \rho_{l_k}\\
		\hat \rho_{l_1} & 1 & \cdots & \hat \rho_{l_{k-1}}\\
		\vdots & \vdots & \ddots& \vdots\\ 
		\hat \rho_{l_k} & \cdots &\hat \rho_{l_1} & 1\\
	\end{pmatrix} \nonumber
\end{equation}
is approximately the identity matrix. Based on this consideration, they proposed to employ the following statistic:
\begin{equation}
	D_{PR}(\bL)=n \left[1- |\hat{\bf{R}}(\bL)|^{1/k}  \right] 
	\label{D_test}
\end{equation}
where $\hat{|\bf{R}|}$ is the determinant of $\hat {\bf{R}} $.
The authors showed that, when the model is correctly specified, $D_{PR}(\bL)$ is asymptotically distributed as a mixture of chi-squared distributions that can be approximated by a Gamma$(a,b)$ distribution, with $a$ and $b$ depending on $k=m$ and on the number $\eta=p+q$ of estimated parameters.
However, \cite{lin_mcleod} highlighted a number of issues concerning both the test statistic, which may not exist, and the Gamma asymptotic approximation. In the same work, they proposed an improved Monte Carlo version of the test which is usually more powerful than the test provided by \cite{pena_rodriguez}.\\ 
Meanwhile, \cite{pena_rodriguez2006} 
proposed a new test statistic, based on the logarithm of $|\hat{\mathbf{R}}(\bL)|$:
\begin{equation}
	\tilde D_{PR}(\bL)=-\frac{n}{k+1}\ \log |\hat{\bf{R}}(\bL)|	\label{Dtilde_test}
\end{equation}
and showed that its distribution is a linear combination of chi-squared random variables that can be approximated by a Gamma$(a,b)$ distribution with 
\[
a=\frac{3\,(k+1)\,(k-2\,\eta)^2}
       {2\,[2\,k\,(2\,k+1)-12\,(k+1)\,\eta]}\qquad\mbox{and}\qquad
b=\frac{3\,(k+1)\,(k-2\,\eta)}
       {2\,k\,(2\,k+1)-12\,(k+1)\,\eta}\qquad ,
\]
where $\eta=p+q$ is the number of estimated parameters.
Unlike $Q_{LB}(\bL)$ and $D_{PR}(\bL)$, $\tilde D_{PR}(\bL)$ is not particularly affected by the value of $k=m$. However, again, the determinant can be negative and, thus, the statistics may not exist.\\
\cite{mahdimcleod} further improved $D_{PR}(\bL)$ by proposing the statistic
\begin{equation}
	\tilde D_{MM}(\bL)=-\frac{3\,n}{2\,k+1} \log |\hat{\bf{R}}(\bL)| 
	\label{DMM_test}
\end{equation}
which, under the null hypothesis, is asymptotically distributed as $\chi_\nu$, with
$$
\nu=\frac{1.5\, k\, (k+1)}{(2\,k+1)-\eta}
$$
and $\eta=p+q$.
To introduce procedures less sensitive to the value of $k=m$, \cite{fishergallagher} suggested new portmanteau tests based on the trace of the autocorrelation matrix. This led to the statistic
\begin{equation}
	Q_{FG}(\bL)=n\,(n+2) \sum_{i=1}^k \frac{k-i+1}{k}\frac{\hat \rho_{l_i}^2}{n-l_i}
	\label{W_test}
\end{equation} 
which can be interpreted as a weighted Ljung-Box statistic, where the weight of the residual autocorrelation is inversely related to $l_i$.
Similarly, a statistic using the matrix of partial autocorrelations, leads to the weighted version of the Monti statistic, $Q_{MW}(\bL)$.\\
The authors proved that under the null hypothesis of an adequately fitted model, $Q_{FG}(\bL)$ and $Q_{MW}(\bL)$ are asymptotically distributed as a mixture of chi-squared distributions, which can be approximated by a Gamma$(a,b)$, with $a$ and $b$ given by
\[
  a = \frac{3}{4}\ \frac{k\, (k+1)^2}{2\, k^2+3\, k+1-6\, k \eta}\qquad\mbox{and}\qquad
  b =\frac{2}{3}\ \frac{2\, k^2+3\, k+1-6\, k\, \eta}{k\, (k+1)}\qquad ,
\]
with $\eta=p+q$.
In a following work, \cite{Gallagher_Fisher_2015} generalized this approach to increase the power of the test by taking general weighted sums of $k$ squared sample autocorrelations:
\begin{equation}
	Q_{WGF}(\bL)=n \sum_{i=1}^k w_i \hat \rho_{l_i}^2
	\label{W2_test}
\end{equation}
using different weighting systems. In particular, they considered
kernel-based weights, geometrically decaying weights, and
data-adaptive weights. According to the authors, allowing the weights
to converge to zero rapidly improves the asymptotic distributional
approximation and increases the power in detecting residual
correlation at low lags.

\textcolor{black}{It should be noted that, unlike the~\cite{ljungbox}
and~\cite{Monti_1994} tests, whose asymptotic null distributions are
generally adequate for practical sample sizes, the other tests
mentioned above rely on considerably more complex asymptotic
distributions whose finite-sample approximation can be
poor. Indeed,~\cite{lin_mcleod} and~\cite{mahdimcleod}
suggest using Monte Carlo calibration to provide improved
finite-sample accuracy.}

\subsection{Portmanteau tests for seasonal residual autocorrelation}

The use of previous tests, with $\bL = (1,\ldots,m)$, in the case of $SARMA(p,q)(P,Q)_S$ models, for example, using $m=4\cdot S$, would be unsuitable and misleading because all lags less than or equal to $m$ would be treated the same, without focusing on the seasonal structure of the autocorrelations.\\
\cite{McLeod_1978} showed that, under the hypothesis of a correctly specified seasonal model and hence of white noise residuals, the residual autocorrelations at lags corresponding to the first $m$ multiples of the period, $S, 2\,S,\ldots, m\,S$, have the same covariance matrix as the first $m$ residual autocorrelations in a nonseasonal model. 
Based on this property, \cite{McLeod_1978} proposed to extend the test statistic $Q_{LB}(\bL)$ defined in~(\ref{LB_test}), with $\bL = (S,2\,S,\ldots,m\,S)$ and proved that, if the model $SARMA(p,q)(P,Q)_S$ has been correctly specified and, thus, residuals are pure white noise, the statistic is asymptotically distributed as $\chi^2_\nu$, with $\nu=m-\eta$ and $\eta=P+Q$.\\
This test is more powerful with respect to residual seasonal autocorrelation. However, note that non-periodic autocorrelation is not tested in this situation.

Exploiting the findings of \cite{McLeod_1978}, \cite{Mahdi_2016} introduced seasonal versions of \cite{pena_rodriguez2006}, \cite{mahdimcleod},  \cite{fishergallagher} and \cite {Gallagher_Fisher_2015} tests. \textcolor{black}{Clearly, the seasonal versions of the test inherit the distributional issues of the original non-seasonal tests.}\\
The test statistics and their asymptotic distributions are as defined above, with $\bL = (S,2\,S,\ldots,$ $m\,S)$ and $\eta=P+Q$.

\section{Generalized portmanteau tests of goodness-of-fit}
\label{sect:generalized}

\textcolor{black}{In the remainder of the paper, we focus on the generalized versions of the~\cite{ljungbox}
and~\cite{Monti_1994} tests. These two tests exhibit complementary performance, in the sense that neither uniformly dominates the other, and the Monti test is generally more effective at detecting omitted MA components. Moreover, as discussed above, their asymptotic null distributions are generally adequate for practical applications.}

\textcolor{black}{As shown above, the other portmanteau tests can also be generalized to simultaneously assess short-range dependence and/or multiple seasonalities. However, an accurate implementation of these tests in practical applications would require Monte Carlo calibration. Including such procedures would considerably complicate the subsequent analysis without substantially contributing to the main objectives of this paper, which are not to compare the relative merits of the various portmanteau tests, but rather to demonstrate that they can be extended to simultaneously assess short-range dependence and/or multiple seasonalities.}

\textcolor{black}{We consider the class of multi-Seasonal ARMA\footnote{The wider class of models proposed in \cite{Lisi_Grigoletto} refers to mSARIMA models, possibly including non-stationary components.} models, recently proposed by \cite{Lisi_Grigoletto} as an extension of the class of SARMA models and explicitly designed to account for multiple seasonality. mSARMA models assume there are $n_s$ periodic components of periods $S_i$ $(i=1,\ldots,n_s)$ and describe the process dynamics as:
\begin{equation}
	\phi(B)\,  \prod_{i=1}^{n_s}\Phi(B^{S_i})\,
	Y_t=\theta(B)\, \prod_{i=1}^{n_s}\Theta(B^{S_i})\, \varepsilon_t,
	\label{msarima}
\end{equation}
where $B$, $\phi(B)$ and $\theta(B)$ are the usual  lag-operator and the autoregressive and moving average
polynomials in $B$ of degrees, respectively, $p$ and $q$;
$\Phi(B^{S_i})=(1- \Phi_{i,1}B^{S_i} - \ldots - \Phi_{i,P_i}B^{P_i
	S_i})$ is the $i$-th seasonal AR polynomial of degree $P_i$ in
$B^{S_i}$; $\Theta(B^{S_i})=(1-\Theta_{i,1}B^{S_i} -
\ldots - \Theta_{i,Q_i}B^{Q_i S_i})$ is the corresponding MA seasonal
polynomial of degree $Q_i$. Polynomials $\phi(B)$ and $\theta(B)$ describe the non-seasonal behavior of the time series, while the polynomials $\Phi(B^{S_i})$ and $\Theta(B^{S_i})$  model the periodic correlation of the $n_s$ seasonal
components of period $S_i$, $(i=1, \ldots,n_s)$.
The model is briefly referred to as an mSARMA$(p,q) \times (P_1, Q_1)_{S_1}\times \ldots\times(P_{n_s},  Q_{n_s})_{S_{n_s}}$.
The values $(p,q)$  and $(P_i, Q_i)$, are the orders of the model, while $S_i$, $i=1, \ldots,n_s$, are the periods of the $n_s$ seasonal components.}\\
\textcolor{black}{Let 
	$\rho({\bf L})$ and $\widehat{\rho}({\bf L})$ be the vectors of the \textcolor{black}{true autocorrelations of the residuals} and of the corresponding estimates computed for} a generic lag vector ${\bf L}=(l_1,\ldots,l_k)$ including any combination of short-term and (multiple) seasonal terms.  This will allow us to consider more complex autocorrelation patterns: a suitable choice of the lags in the vector ${\bf L}$ permits to test the absence of multi-seasonal residual autocorrelation as well as the joint absence of short-term and (multi-)seasonal residual linear dependence. \textcolor{black}{In Section \ref{sect:multi}}, these two situations will be described in detail.

\textcolor{black}{The following theorem generalizes the results of \cite{McLeod_1978} and allows us to derive the asymptotic distribution of the test statistics in the multiseasonal framework.}\\

\noindent {\bf Theorem 1.}
	\textcolor{black}{{\it Assume that the stationary, invertible, non-redundant mSARMA$(p,q)\times(P_1,Q_1)_{S_1}$ $\times\cdots\times(P_{n_s},Q_{n_s})_{S_{n_s}}$
	model is correctly specified and that the parameter estimators are asymptotically efficient (such as ML estimators). Then, for any fixed lag vector ${\bf L}=(l_1,\dots,l_k)$,
	$$ 
	\sqrt{n}\,\widehat{\rho}({\bf L})
	\overset{d}{\longrightarrow}
	\mathcal{N}\!\left(
	\mathbf 0,\;\mathbf I_k- \mathbf X_L \Fisher^{-1} \mathbf X_L^T
	\right),
	$$
	where $\mathbf I_k$ is the $k\times k$ identity matrix,
        $\Fisher$ is the per-observation information matrix of the complete
        mSARMA model and $\mathbf X_L$ is \textcolor{black}{the matrix whose rows are defined as in (A.2)}.}}\\
	\hspace{1cm}\\
	\textcolor{black}{The proof, given in Appendix, follows the three-lemma scheme used in \cite{McLeod_1978} with the main new step being the computation of the score derivatives $\partial \hat \varepsilon_t/\partial \beta_i$ when the autoregressive and moving-average operators are \emph{products} of several
	seasonal factors rather than a single polynomial. \\
	The above theorem shows that the asymptotic distribution of
        $\widehat{\rho}(\bf L)$ depends only on the number of selected lags and on the estimation effect introduced through the projection matrix
	$\mathbf X_L \Fisher^{-1} \mathbf X_L^T$.
	Consequently, the specific location of the selected lags does not modify the structure of the asymptotic result; only the covariance correction due to parameter estimation affects the limiting distribution. This extends the argument originally proposed by \cite{McLeod_1978} for multiplicative seasonal ARMA models to the broader class of multi-seasonal models.} \\
	\hspace{1cm}\\
\noindent {\bf Corollary 1.}	\label{coroll}
\textcolor{black}{{\it Suppose $m_0,m_1,\dots,m_{n_s}$ are large enough that the impulse function
coefficients are negligible beyond the corresponding lags, and that the roots of
$\phi(z)\theta(z)\prod_i\Phi_i(z^{S_i})\Theta_i(z^{S_i})=0$ are not close to the unit
circle. Then $\mathbf X_L^{T} \mathbf X_L \doteq \Fisher$, so that the asymptotic covariance matrix of $\sqrt{n}\,\hat{\rho}(\bf L)$ is idempotent of rank $k-\eta$, and the generalized
Ljung--Box statistic
\[
Q_{LB}({\bf L}) = n(n+2)\sum_{i=1}^{k}\frac{\hat \rho^2(l_i)}{n-l_i}
\]
converges in distribution to $\chi^2_{k-\eta}$, with
\[
\eta = p+q+\sum_{i=1}^{n_s}(P_i+Q_i)\ .
\]
}}
\textcolor{black}{The same argument applied to the partial residual autocorrelations gives the analogous
asymptotic $\chi^2_{k-\eta}$ distribution for the generalized Monti statistic (G-M).}\\
Details on how to adapt these generic expressions in two specific situations will be given in the following section.\\ 

\subsection{Portmanteau tests for multi-seasonal residual \textcolor{black}{and mixed}  autocorrelation}
\label{sect:multi}

\textcolor{black}{In this section we consider two important issues that can be faced using generalized portmanteau tests.}\\
The use of high-frequency data, particularly at intra-day frequencies, is increasingly widespread. This leads to the increasingly frequent presence of multiple periodicities in time series, which must be handled and tested appropriately. In particular, with respect to the problem considered in this article, this requires the availability of tests that can jointly assess the absence of residual seasonal correlation of multiple cyclical components. \textcolor{black}{To the best of our knowledge}, portmanteau tests specifically built to analyze the residual autocorrelation of multi-seasonal time series are not yet present in the literature.\\
\textcolor{black}{In the context of the mSARIMA models,} we want to verify that the residual time series of an estimated mSARMA model does not show periodic autocorrelation for all the $n_s$ periods.
With respect to the system of hypotheses (\ref{sys_hyp}), this can be done by choosing the lag vector 
\[
\bL=\left(\{S_1,\ldots,m_1\, S_1\} \cup \ldots \cup \{S_{n_s},\ldots,m_{n_s}\, S_{n_s}\}\right)\ ,
\]
considering all seasonal lags and their $(m_1, m_2,\ldots, m_{n_s})$ multiples. Clearly, some of these lags can overlap; thus, the dimension $k$ of the lag vector $\bL$ is generally smaller than $m_1 +\ldots+ m_{n_s}$.\\
We suggest to use the test statistics given above, with this vector $\bL$.\\
Under $H_0$ and asymptotically, the statistics follow the given distributions, with number of estimated parameters, $\eta=\sum_{i=1}^{n_s}  (P_{i} + Q_{i})$.\\
\noindent Another issue that has not yet received \textcolor{black}{sufficient} attention in the literature  is the possibility of jointly \textcolor{black}{testing} the presence of short-term and (multiple) seasonal residual autocorrelation. An example of system of hypotheses to test, in the case of a single seasonal component of period $S$ is system~(\ref{sys_hyp}), with
\[
\bL = \left(1,\ldots,m_0,S,\ldots,m_1\,S\right)\ \,
\]
where the absence of residual autocorrelation is tested on the first $m_0$ consecutive lags and on the first $m_1$ seasonal lags.\\
To our knowledge, no formal statistical test has yet been introduced to
verify such a system of hypotheses. As a consequence, to test the
absence of short-term and seasonal residual autocorrelation one should
consider two separate tests, one for short-term autocorrelation and a
second one for seasonal residual effects. Although this can provide
some insights on the residual linear dependence structure, this poses
the problem of the global level of the test. This is a tricky point
both because it needs to combine the levels of two, possibly
dependent, tests and because even when the null hypothesis of each test is
true, the level of this test is no longer correct, since when
there is no short-term correlation, seasonal correlation remains, and
vice versa.

As an example, Table~\ref{test_misto} lists the empirical levels ($\alpha_{obs}$), computed on $N=2000$ Monte Carlo replications, of the portmanteau tests of seasonal residual autocorrelation.
The series, of length $n=500$ are generated from a SARMA$(p,q)(P,Q)_4$ process with different orders and values of the parameters. The fitted models correctly identify the seasonal part but neglect the non-seasonal component. Note that, for the residuals of the fitted models, the null hypothesis is true or approximately true. Indeed, for MA$(1)$ processes, $\rho_k=0$ for $k>1$, for an AR(1) with $\phi=0.4$ we have  $\rho_k<0.026$ for $k\geq 4$ and for an ARMA(1,1) with $\phi=0.4$ and $\theta=0.4$ we have $\rho_k<0.045$ for $k\geq 4$.
In all cases listed in Table~\ref{test_misto}, an exact binomial test of equivalence between the observed and the nominal level is significant. Thus, despite the null hypothesis being true, \textcolor{black}{deviations} from the white noise assumption \textcolor{black}{cause} the tests to over-reject $H_0$.\\

\begin{table}[H]
\centering
\begin{small}
 \begin{tabular}{cccccc}
   \hline
   $\phi_1$ & $\theta_1$ & $\Phi_1$ & $\Theta_1$ & LB & M  \\ \hline
  \multicolumn{6}{l}{Fitted model: SAR(1)} \\ \hline
  *& 0.5& 0.5& *& 0.137 & 0.158  \\ 
  0.4 & *& 0.5& *& 0.153 & 0.125  \\ 
  0.4 & 0.4 & 0.5& *& 0.316 & 0.356  \\ \hline
\multicolumn{6}{l}{Fitted model: SMA(1)} \\ \hline
  *& 0.5& *& 0.5& 0.129 & 0.160  \\ 
  0.4 & *& *& 0.5& 0.146 & 0.132  \\ 
  0.4 & 0.4 & *& 0.5& 0.309 & 0.347  \\ 
   \hline
\end{tabular}
\caption{Empirical level ($\alpha_{obs}$) of a seasonal portmanteau test at nominal level $5\%$. 
LB=Ljung-Box test, M=Monti test. $N=2000$, $n=500$.} 
\label{test_misto}
\end{small}
\end{table}
The issues discussed above make it difficult to have global tests with well-defined levels. Instead, a suitable generalized portmanteau test for the system of hypotheses~(\ref{sys_hyp}) with a lag vector $\bL=(1,\ldots,m_0,S,\ldots,m_1\,S)$, i.e.\ including the first $m_0$ non-seasonal lags and the first $m_1$ seasonal lags, allows one to correctly address the problem \textcolor{black}{using a single test that maintains the nominal significance level}.

\section{Simulation study}

In this section, we explore the performance of the proposed statistics using Monte Carlo simulations. 
In particular, we analyze the empirical features of the generalized
versions of the non-seasonal and \textcolor{black}{seasonal~\cite{ljungbox}
and~\cite{Monti_1994} portmanteau tests (G-LB and G-M henceforth)} focusing on two cases not
covered by the existing literature, i.e.\ multi-seasonal and mixed
(short-term and seasonal) residual autocorrelation.\\
Unlike most studies in the literature, the analyses are not limited to the computation of the observed level and power of the tests, but the equivalence between observed and nominal levels is assessed using an exact two-sided binomial test. Similarly, the agreement between the Monte Carlo distributions of the test statistics and the expected ones is formally assessed by applying the Kolmogorov-Smirnov and Anderson-Darling tests.\\ 
\textcolor{black}{Across all simulations, two pairs of periodicities were considered: $(S_1,S_2)=(4,7)$, representing non-overlapping cycles, and $(S_1,S_2)=(4,12)$, as an example of overlapping cycles. Since the results for these two cases were substantially similar, for the sake of brevity, only the results for $(S_1,S_2)=(4,7)$ are reported in most of the analyses. However, the results for $(S_1,S_2)=(4,12)$ are available from the authors upon request. }

\subsection{Sensitivity to $m$ and $n$}
\label{sect:dependence}

\begin{figure}[H]
 	\begin{center}
 	\includegraphics[height=4.8cm, width = 4.8cm]{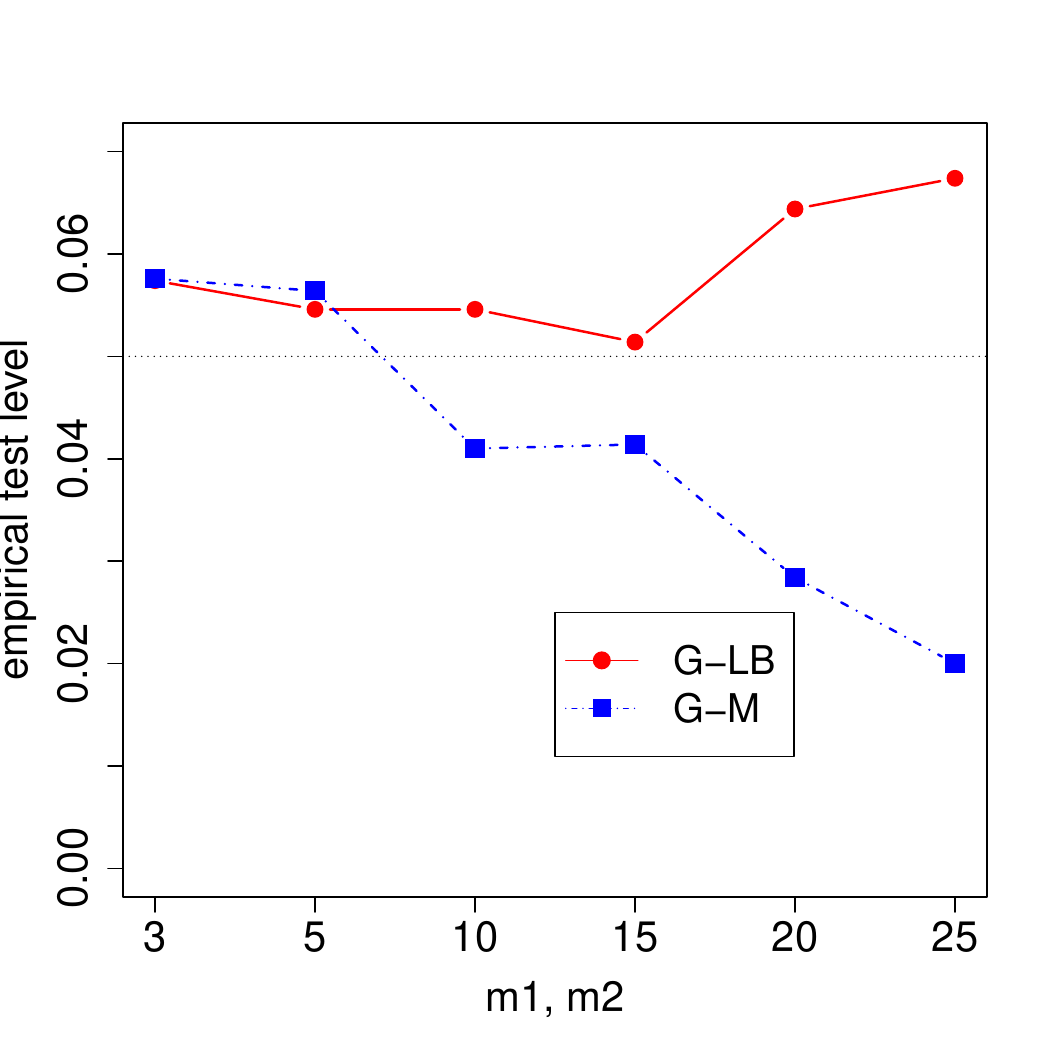}
        \includegraphics[height=4.8cm, width = 4.8cm]{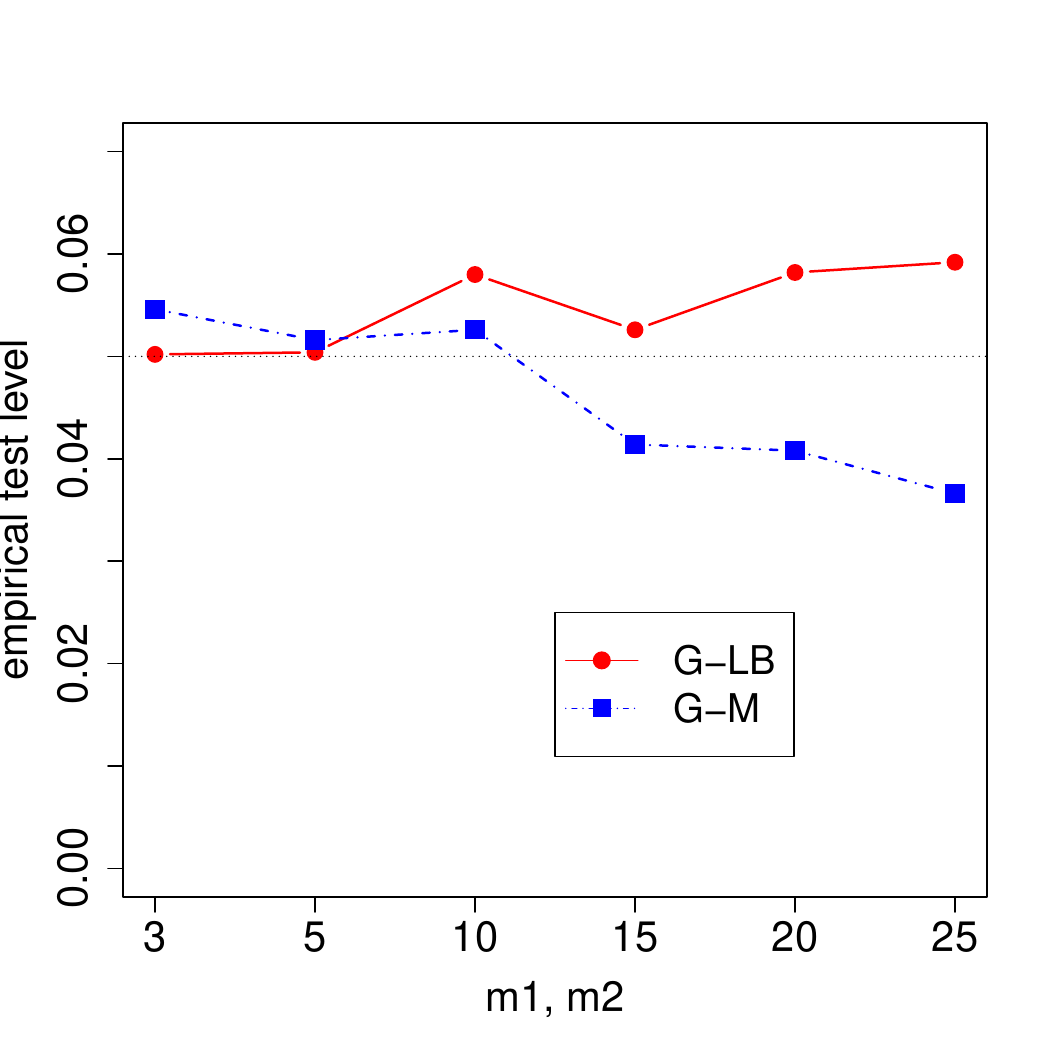}
        \includegraphics[height=4.8cm, width = 4.8cm]{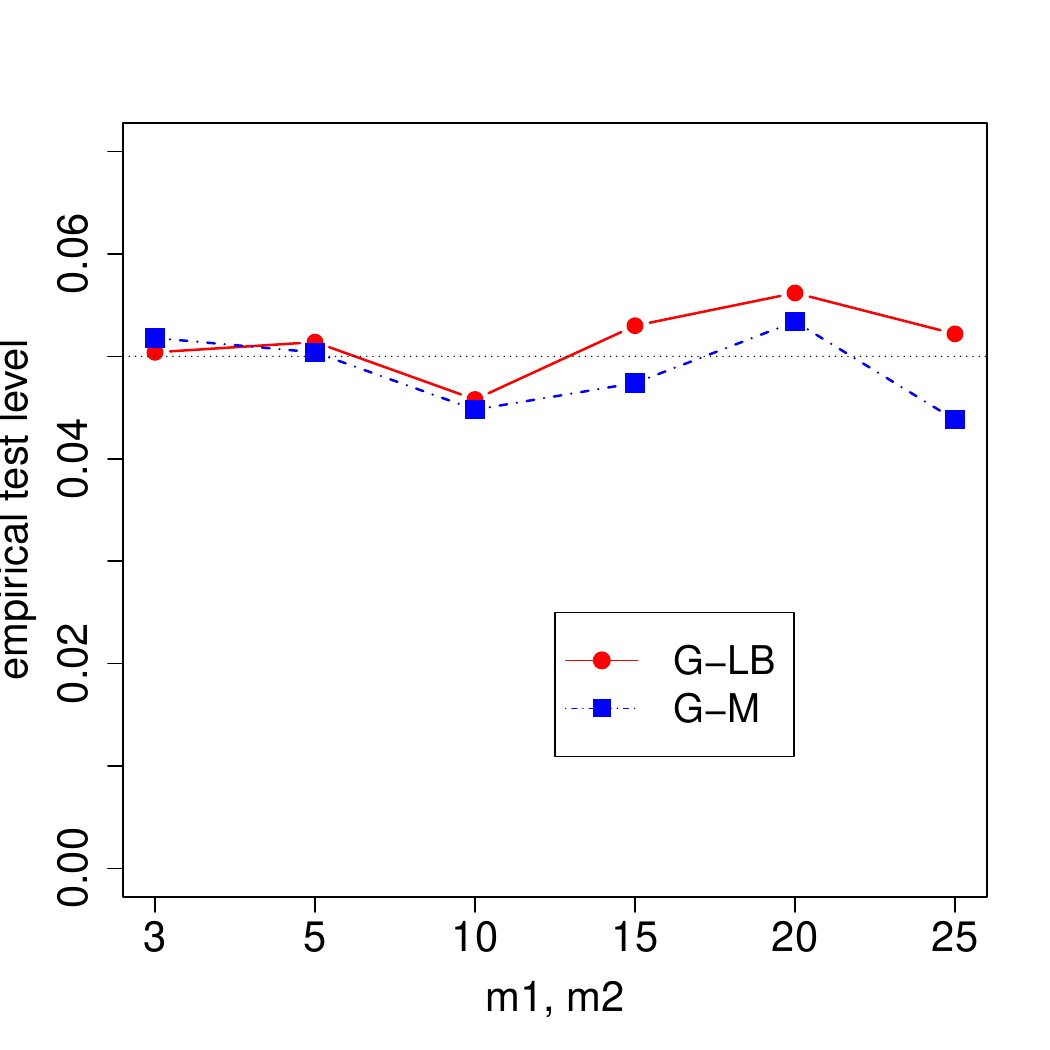}
        \\
 	\includegraphics[height=4.8cm, width = 4.8cm]{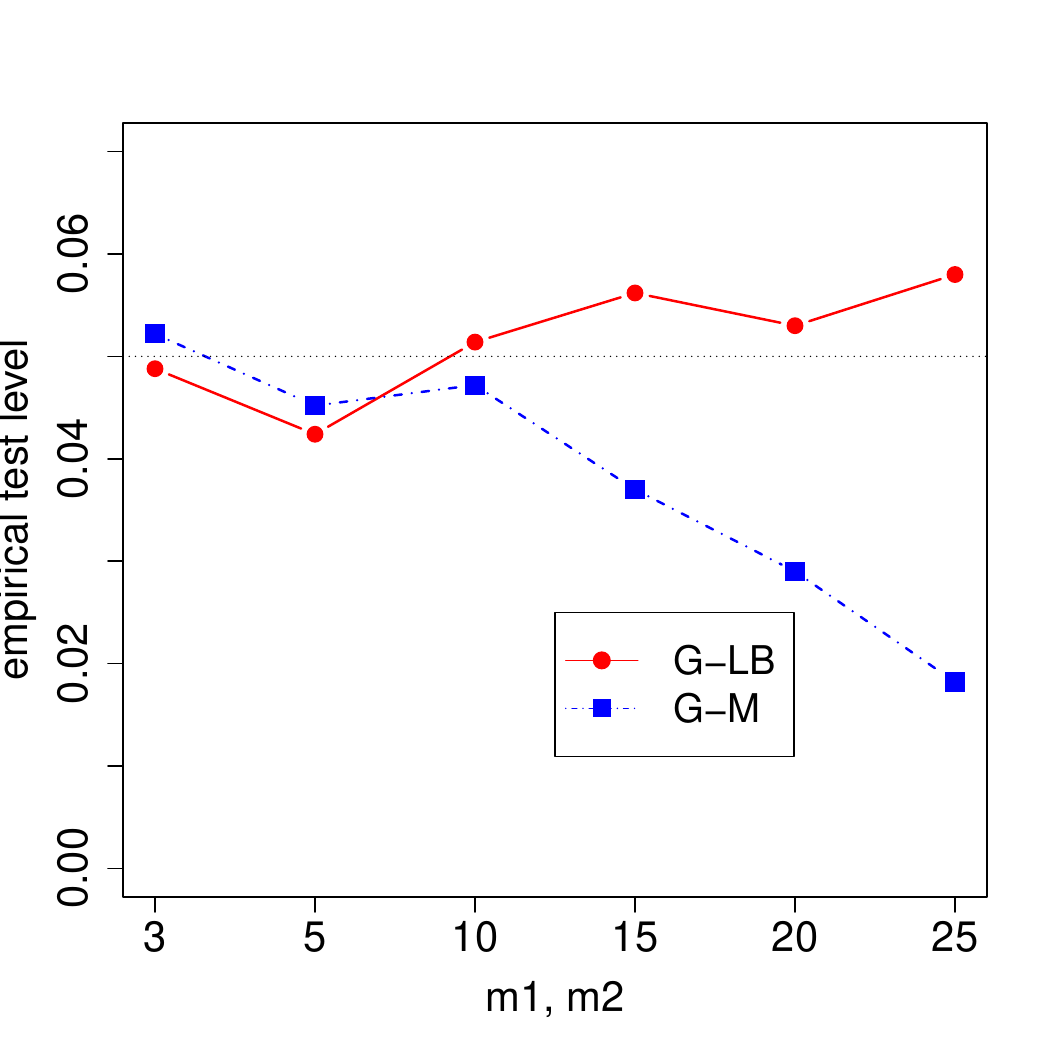}
        \includegraphics[height=4.8cm, width = 4.8cm]{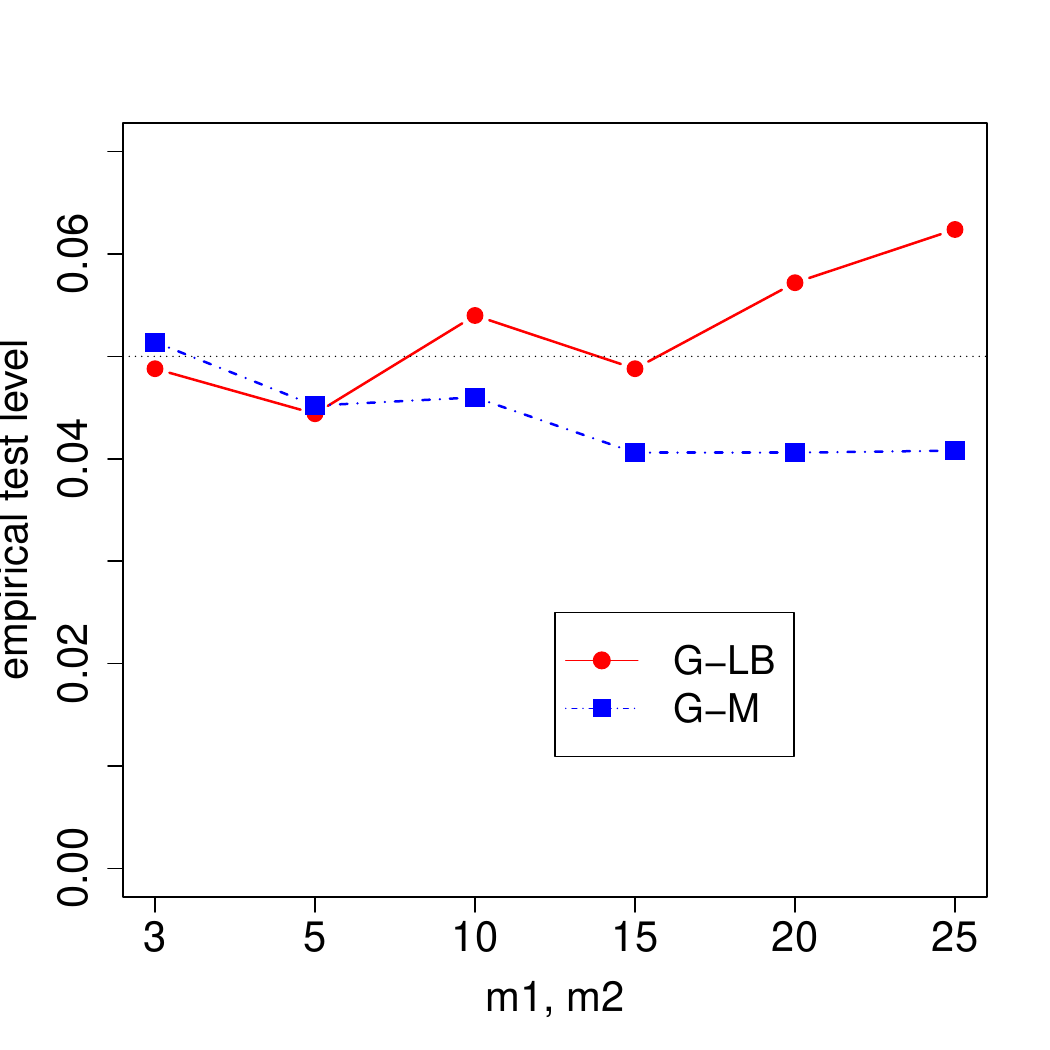}
        \includegraphics[height=4.8cm, width = 4.8cm]{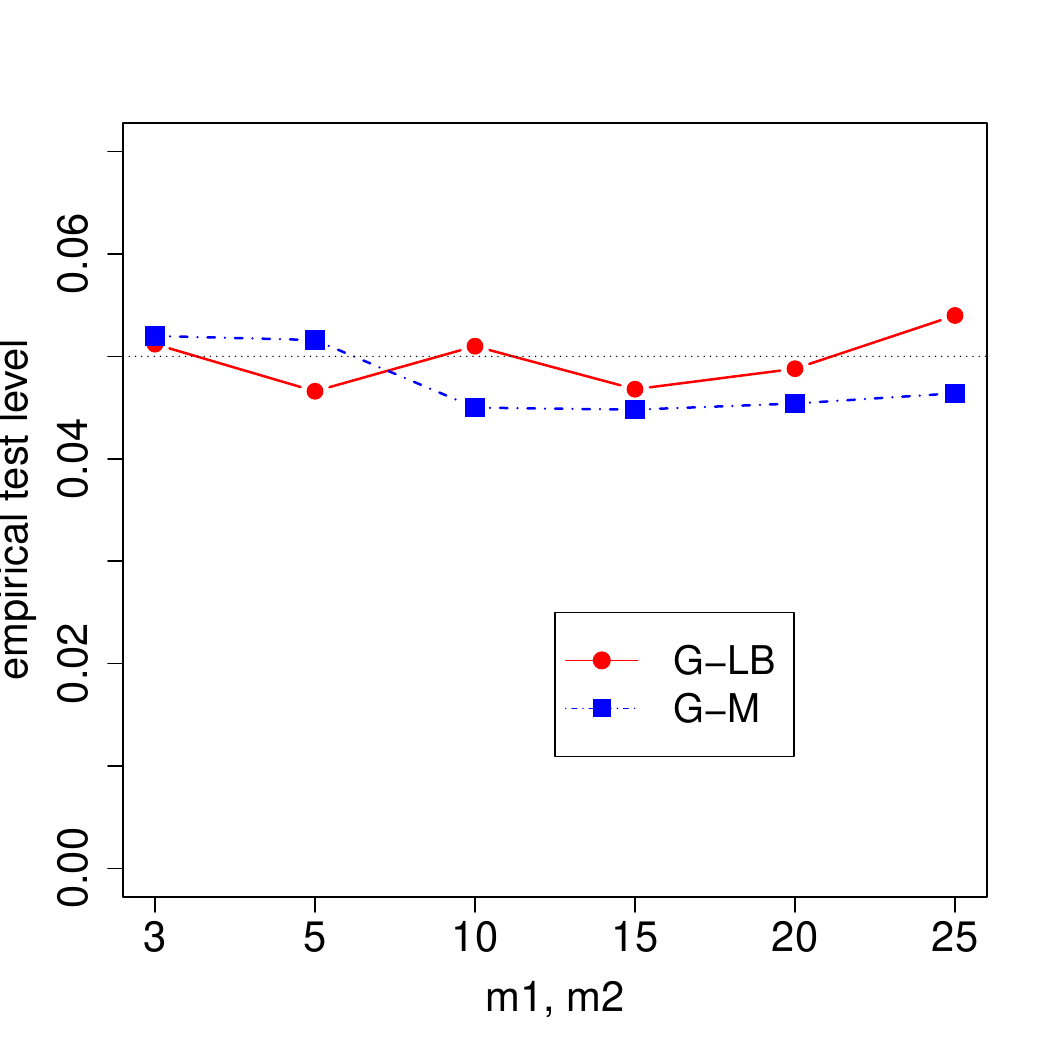}
        \\
        \includegraphics[height=4.8cm, width = 4.8cm]{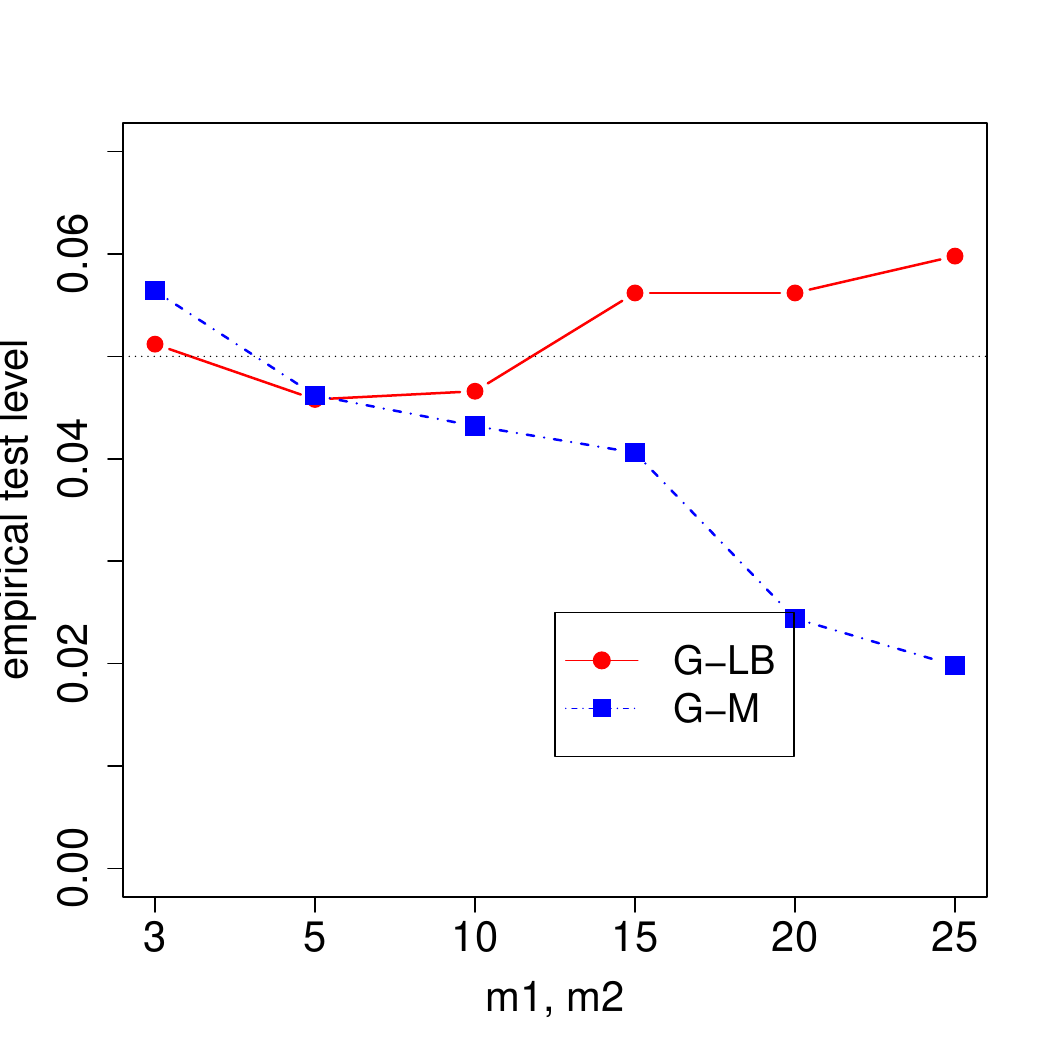}
        \includegraphics[height=4.8cm, width = 4.8cm]{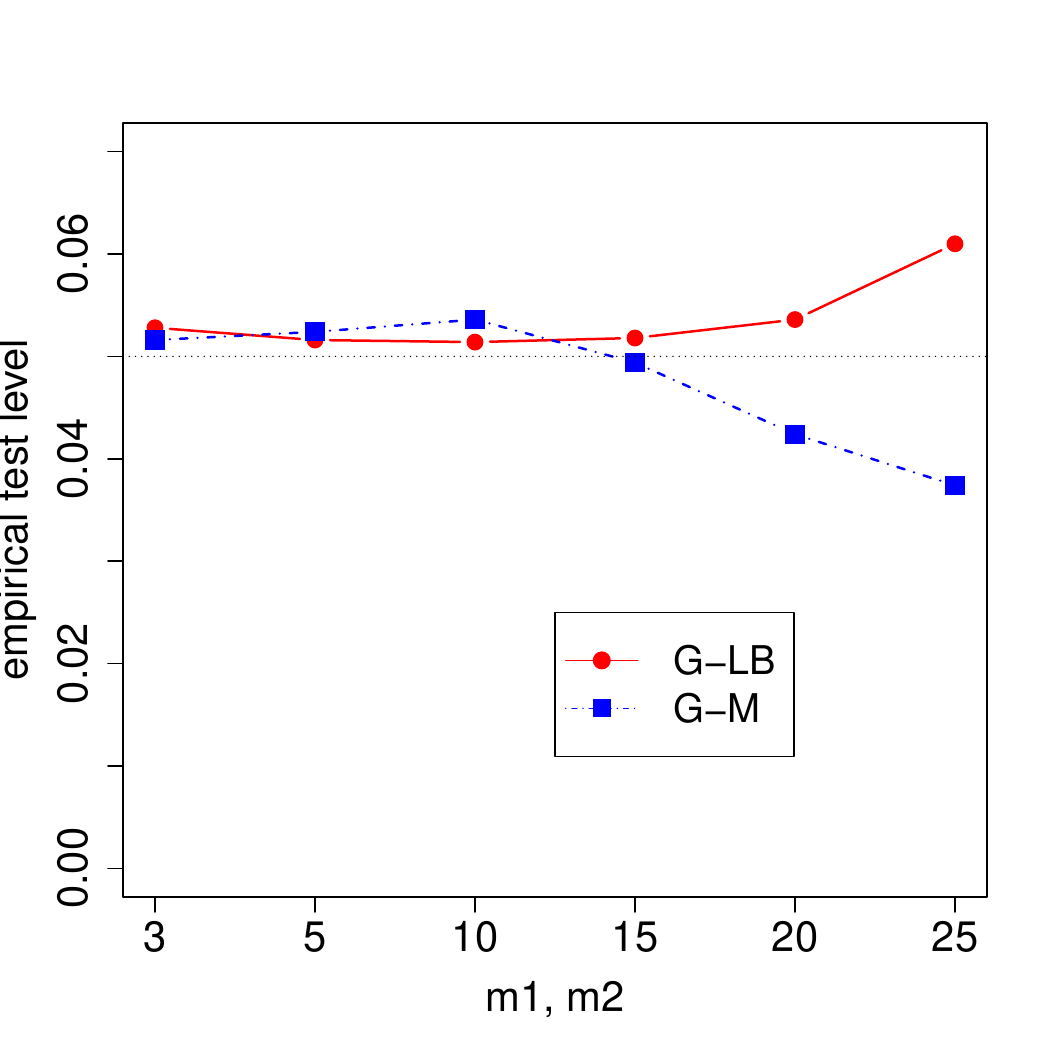}
        \includegraphics[height=4.8cm, width = 4.8cm]{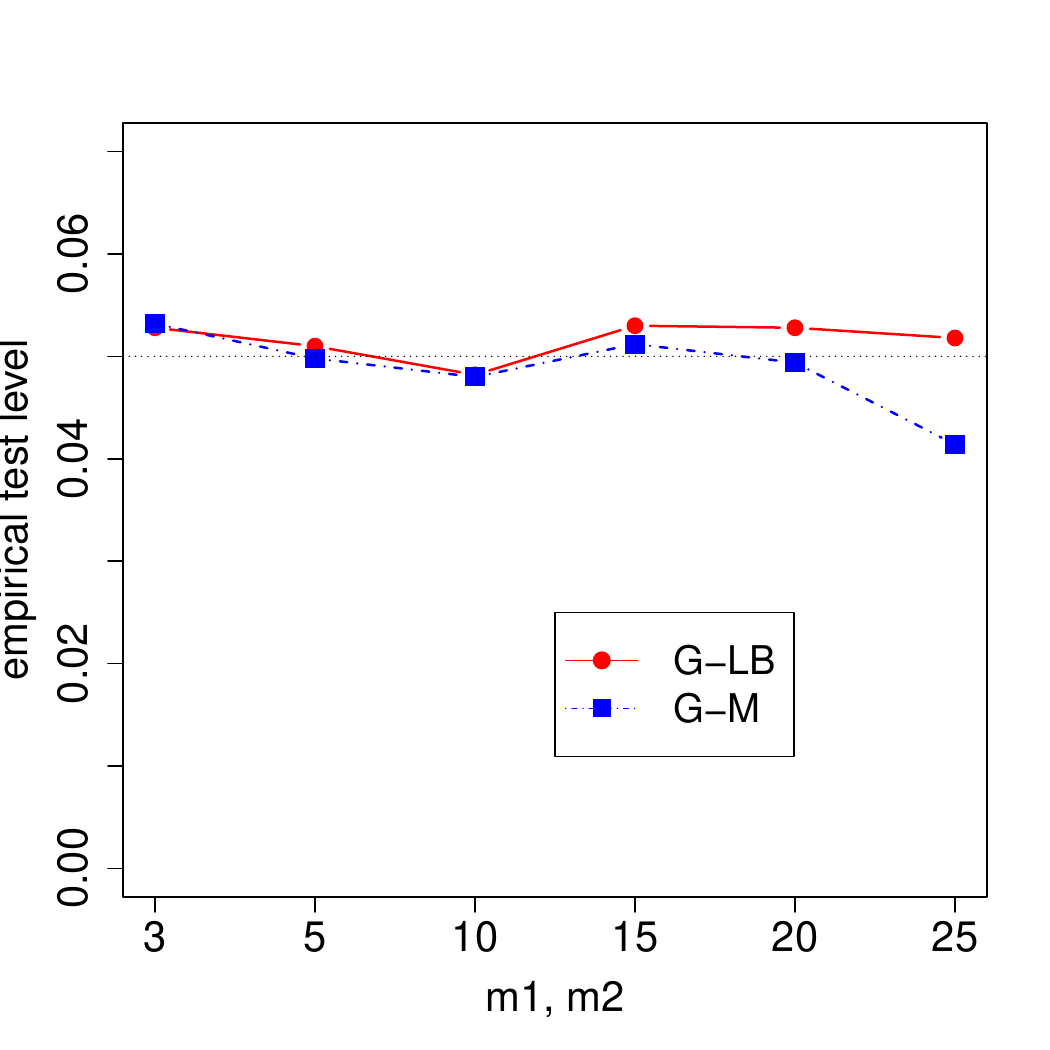}
        \\
 		\caption{Empirical levels as functions of $m_1$ and $m_2$.
        \textcolor{black}{Columns 1, 2 and 3 refer to $n=250$, $n=500$ and $n=1000$, respectively.
        DGP: double SAR(1) with $\Phi_{1,1}=0.5$ and $\Phi_{2,1}=0.4$ for the first row; double SMA(1) with $\Theta_{1,1}=0.5$ and $\Theta_{2,1}=0.4$ for the second row; mSARMA(0,0,0)(1,0,0)$_4$(0,0,1)$_7$ with $\Phi_{1,1}=0.5$ and $\Theta_{2,1}=0.4$.
        In all cases $s_1=4$ and $s_2=7$. The dotted horizontal line is the nominal level 0.05. Red circles with continuous line: G-LB; blue squares with dashed line: G-M.}}
 		\label{dep_on_m}
 	\end{center}
\end{figure}

\begin{figure}[H]
 	\begin{center}
 		\includegraphics[height=4.8cm, width=4.8cm]{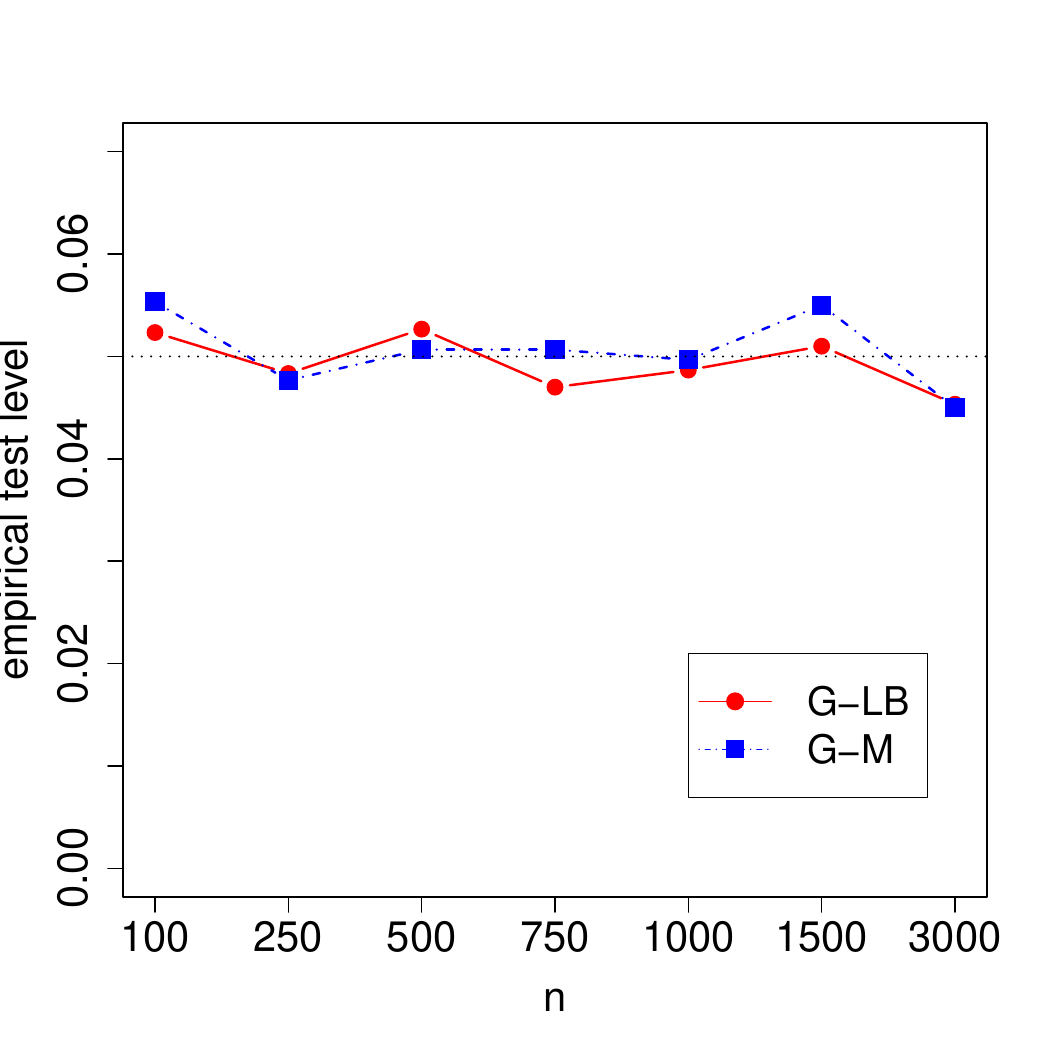}
         		\includegraphics[height=4.8cm, width = 4.8cm]{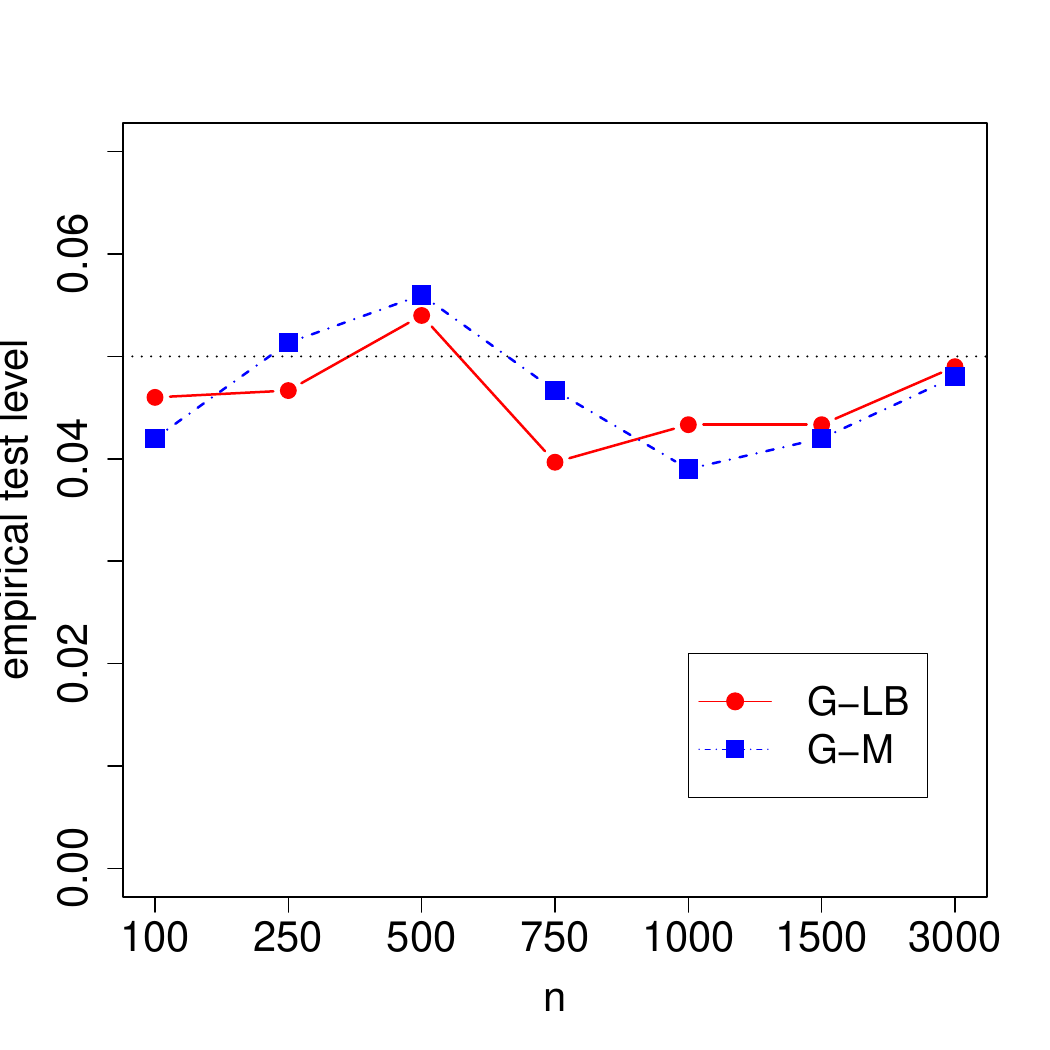}
 		\includegraphics[height=4.8cm, width=4.8cm]{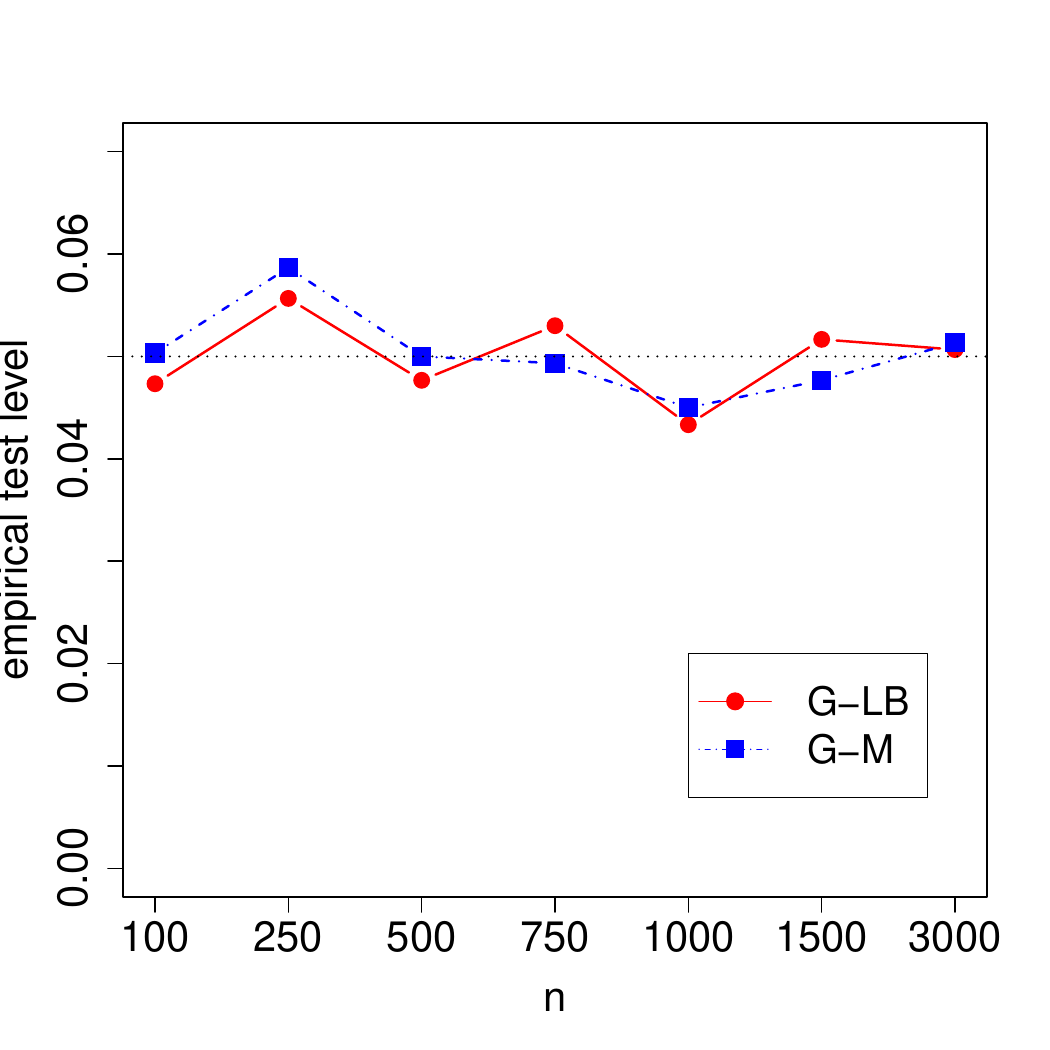}
 		\caption{Empirical levels as a function of $n$.
        \textcolor{black}{First panel: mSARMA(0,0,0)(1,0,0)$_4$(1,0,0)$_7$ with $\Phi_{1,1}=0.5$ and $\Phi_{2,1}=0.4$; second panel: mSARMA(0,0,0)(0,0,1)$_4$(0,0,1)$_7$ with 
        $\Theta_{1,1}=0.5$ and $\Theta_{2,1}=0.4$; 
    	third panel: mSARMA(0,0,0)(1,0,0)$_4$(0,0,1)$_7$ with $\Phi_{1,1}=0.5$ and $\Theta_{2,1}=0.5$. In all cases $m_1=m_2=5$. Red circles with continuous line: G-LB; blue squares with dashed line: G-M.}}
        \label{dep_on_n}
 	\end{center}
\end{figure}

As a first, preliminary step, we analyze the sensitivity of the tests to $m$ and $n$.\\
The choice of the number of lags, $m$, is a debated topic in the portmanteau testing literature, and, today, there is no universally accepted solution.
Indeed, the choice of $m$ represents a classic trade-off: for small values of $m$ the test may have low power against alternatives with autocorrelation at high lags, but is statistically more stable; for large values of $m$ the test is more sensitive to autocorrelation at high lags, but tends to lose power because many near-zero terms are added under the alternative hypothesis, effectively diluting the signal.\\
However, none of these rules has been shown to be universally superior \textcolor{black}{and the best choice for $m$ relies on the time series’ length as well as the test’s level \citep{Hassani_Yeganegi_2020}}.\\  
Some practical advice  or rules of thumb have been proposed in the literature: \textcolor{black}{
$m=5$, \citep{ljung}, $m=20$ \citep{Shumway_Stoffer_2011},  $m=\ln(n)$ \citep{Tsay2010}, $m=\min\{ 10, n/5 \}$ \citep{Hyndman_Athanasopoulos_2018}}. 
In addition, \cite{Escanciano_2009} proposed an automatic selection procedure that employs a data-driven penalty function based on both AIC and BIC to automatically select the optimal number of lags. \\
Here, we don't want to discuss the optimal choice of $m$ but, more simply, to show how sensitive the tests are to the values of $m$ and $n$. This allows us to find a value of $m$, to be used in the Monte Carlo simulations, for which the tests' performance is good and stable enough. \\
%
Figure~\ref{dep_on_m} shows the empirical level of the two tests as a function of $m_1$ and $m_2$ with $m_1=m_2$ and for $n=250$, $500$ and $n=1000$.
The panels in the first row depict the empirical levels for a double SAR(1) with $\Phi_{1,1}=0.5$ and $\Phi_{2,1}=0.4$, those in the second row refer to a double SMA(1) with $\Theta_{1,1}=0.5$ and $\Theta_{2,1}=0.4$ and those in the third row are related to a mSARIMA$(0,0,0)(1,0,0)_{s_1}(0,0,1)_{s_2}$ with $\Phi_{1,1}=0.5$ and $\Theta_{2,1}=0.4$. In all cases $s_1=4$ and $s_2=7$ and the values of $m_i$ range between 3 and 25. The graphs suggest that both tests work better for small values of $m$. In addition, the G-LB test appears to be less sensitive to the choice of $m_i$ for relatively small values of $n$ even if for higher values of $n$ both of them tend to behave similarly. These results are in line with the literature which suggests using small values of $m$ for \citep{ljung}.\\
Globally, these first findings suggest that choosing \textcolor{black}{about} $m=5$ lags for each component is a good choice for both tests. Also, five lags are sufficient to detect possible residual dependence without introducing so many lags that the signal becomes diluted.\\
As concerns the dependence on the length of the series, Figure~\ref{dep_on_n} exhibits the empirical level for different sample sizes and $m_1=m_2=5$, for a double SAR(1), a double SMA(1) and a mixed SAR/SMA data-generating process, for $(s_1,s_2)=(4,7)$. \\
Both for the sensitivity on $m$ and $n$, the same analysis was conducted for $(s_1,s_2)=(4,12)$ but, since the results are basically the same, they are not reported.\\
Overall, \textcolor{black}{these findings suggest that the performance of the tests is not dramatically affected by sample size} and, starting from $n=100$ the generalized version of the tests leads to relatively stable results with respect to $n$.

\subsection{Portmanteau test for multi-seasonal residual autocorrelation}
\label{sect:multiseas}

We now analyze the generalized versions of the~\cite{ljungbox} and~\cite{Monti_1994} tests.\\ 
Again, the empirical properties of the test are based on $N=2000$ simulated time series of length n = $250, 500$ and $1000$. 
In the multi-seasonal case we consider also the sample size $n=1000$ because, in this context, it is quite common to have thousands of data. \textcolor{black}{For example, with hourly data, one thousand observations correspond to just over 42 days}. \\
We generate series from mSARIMA$(p,d,q)(P_1,D_1,Q_1)_{S_1}(P_2,D_2,Q_2)_{S_2}$ models \citep{Lisi_Grigoletto}, for different orders and values of parameters. In particular, we considered $(S_1,S_2)$ equal to $(4,7)$ and $(4, 12)$ as representative of non-overlapping and overlapping periodic lags.\\
To explore the multiple seasonal case, we consider time series with two seasonal components of periods $(S_1,S_2)$, without non-periodic components, and we apply the generalized tests previously described. 
In more detail, to investigate the type-I error of the tests we simulate time series from mSARIMA$(0,0,0)(1,0,0)_{S_1}(1,0,0)_{S_2}$ and mSARIMA$(0,0,0)(0,0,1)_{S_1}(0,0,1)_{S_2}$ models, with $(S_1,S_2)=(4,7), (4,12)$ and for the combinations of parameters listed in Tables~\ref{multis_level_sarsar} and~\ref{multis_level_smasma}. For each time series a correctly identified model is estimated and the portmanteau tests are applied to the residuals.\\
The results \textcolor{black}{for $(S_1,S_2)=(4,7)$} are summarized in Tables~\ref{multis_level_sarsar} and~\ref{multis_level_smasma}, listing the empirical level of the tests when data are generated by a double SAR and a double SMA process, respectively. 
\textcolor{black}{The corresponding tables for $(S_1,S_2)=(4,12)$ are provided in Appendix B.}
In the tables, a single asterisk indicates that $\alpha_{obs}$ is significantly different from the nominal $\alpha$ at a significance level of $5\%$, while the double asterisk means significance at the level of $1\%$.\\
\textcolor{black}{The G-LB and G-M tests exhibit rejection rates in line with the nominal levels}. To confirm this, according to the binomial test, the hypothesis of an equal observed and nominal level is almost never rejected across models and sample sizes.
\begin{table}[H]
	\centering
    \begin{small}
	\begin{tabular}{cc|cc|cc|cc}
		\hline
$\Phi_{1,1}$ & $\Phi_{2,1}$ & G-LB & G-M  & G-LB & G-M  & G-LB & G-M  \\ \hline
      \multicolumn{2}{l}{$\alpha=0.01$}    & \multicolumn{2}{c}{$n=250$}  & \multicolumn{2}{c}{$n=500$} & \multicolumn{2}{c}{$n=1000$}  \\ \hline 
0.200 & 0.200 & 0.007 & 0.011 & 0.014 & 0.012 & 0.006 & 0.006 \\ 
  0.500 & 0.200 & 0.010 & 0.006 & 0.014 & 0.012 & 0.009 & 0.012 \\ 
  0.800 & 0.200 & 0.014 & 0.008 & 0.009 & 0.008 & 0.011 & 0.012 \\ 
  0.200 & 0.500 & 0.007 & 0.008 & 0.014 & 0.013 & 0.009 & 0.010 \\ 
  0.500 & 0.500 & 0.010 & 0.006 & 0.011 & 0.011 & 0.007 & 0.009 \\ 
  0.800 & 0.500 & 0.010 & 0.010 & 0.009 & 0.007 & 0.012 & 0.012 \\ 
  0.200 & 0.800 & 0.010 & 0.009 & 0.013 & 0.012 & 0.010 & 0.013 \\ 
  0.500 & 0.800 & 0.012 & 0.010 & 0.012 & 0.012 & 0.014 & 0.011 \\ 
  0.800 & 0.800 & 0.008 & 0.010 & 0.016$^{*}$ & 0.012 & 0.011 & 0.010 \\ 
 \hline
      \multicolumn{2}{l}{$\alpha=0.05$}    & \multicolumn{2}{c}{$n=250$}  & \multicolumn{2}{c}{$n=500$} & \multicolumn{2}{c}{$n=1000$}  \\ \hline 
  0.200 & 0.200 & 0.050 & 0.054 & 0.052 & 0.052 & 0.046 & 0.044 \\   0.500 & 0.200 & 0.048 & 0.048 & 0.054 & 0.054 & 0.052 & 0.051 \\ 
  0.800 & 0.200 & 0.054 & 0.054 & 0.043 & 0.045 & 0.052 & 0.054 \\ 
  0.200 & 0.500 & 0.038 & 0.045 & 0.049 & 0.051 & 0.048 & 0.050 \\ 
  0.500 & 0.500 & 0.046 & 0.043 & 0.053 & 0.050 & 0.044 & 0.048 \\ 
  0.800 & 0.500 & 0.041 & 0.040 & 0.052 & 0.050 & 0.055 & 0.054 \\ 
  0.200 & 0.800 & 0.050 & 0.058 & 0.059 & 0.049 & 0.046 & 0.047 \\ 
  0.500 & 0.800 & 0.054 & 0.046 & 0.045 & 0.047 & 0.048 & 0.045 \\ 
  0.800 & 0.800 & 0.046 & 0.052 & 0.057 & 0.054 & 0.047 & 0.046 \\ 
    \hline
      \multicolumn{2}{l}{$\alpha=0.10$}    & \multicolumn{2}{c}{$n=250$}  & \multicolumn{2}{c}{$n=500$} & \multicolumn{2}{c}{$n=1000$}  \\ \hline 
 0.200 & 0.200 & 0.100 & 0.106 & 0.102 & 0.108 & 0.090 & 0.084 \\
 0.500 & 0.200 & 0.096 & 0.096 & 0.103 & 0.113 & 0.103 & 0.102 \\ 
  0.800 & 0.200 & 0.096 & 0.106 & 0.091 & 0.100 & 0.105 & 0.108 \\ 
  0.200 & 0.500 & 0.090 & 0.090 & 0.100 & 0.098 & 0.096 & 0.094 \\ 
  0.500 & 0.500 & 0.088 & 0.094 & 0.098 & 0.097 & 0.101 & 0.102 \\ 
  0.800 & 0.500 & 0.078$^{**}$ & 0.088 & 0.106 & 0.099 & 0.108 & 0.106 \\ 
  0.200 & 0.800 & 0.092 & 0.094 & 0.108 & 0.108 & 0.101 & 0.105 \\ 
  0.500 & 0.800 & 0.097 & 0.103 & 0.099 & 0.100 & 0.096 & 0.094 \\ 
  0.800 & 0.800 & 0.101 & 0.096 & 0.100 & 0.102 & 0.090 & 0.088 \\ 
   \hline
	\end{tabular}
    	\caption{Multi-seasonal case: empirical level ($\alpha_{obs}$) of the test. Data generated from an mSARIMA$(0,0,0)(1,0,0)_4 (1,0,0)_7$ model for different values of parameters $\Phi_{1,1}$ and $\Phi_{2,1}$ and sample sizes. G-LB=generalized Ljung-Box test, G-M=generalized Monti test, $\alpha$=nominal test level, $N=2000$, $\bL=(4, 7, 8, 12, 14, 16, 20, 21, 28, 35)$, $(S_1,S_2)=(4,7)$.    	$^*$  and $^{**}$ denote empirical levels significantly different from the nominal one at $5\%$ and at $1\%$, respectively.}  \label{multis_level_sarsar}
        \end{small}
\end{table}

\begin{figure}[H]
 	\begin{center}
 		\includegraphics[width = 5cm]{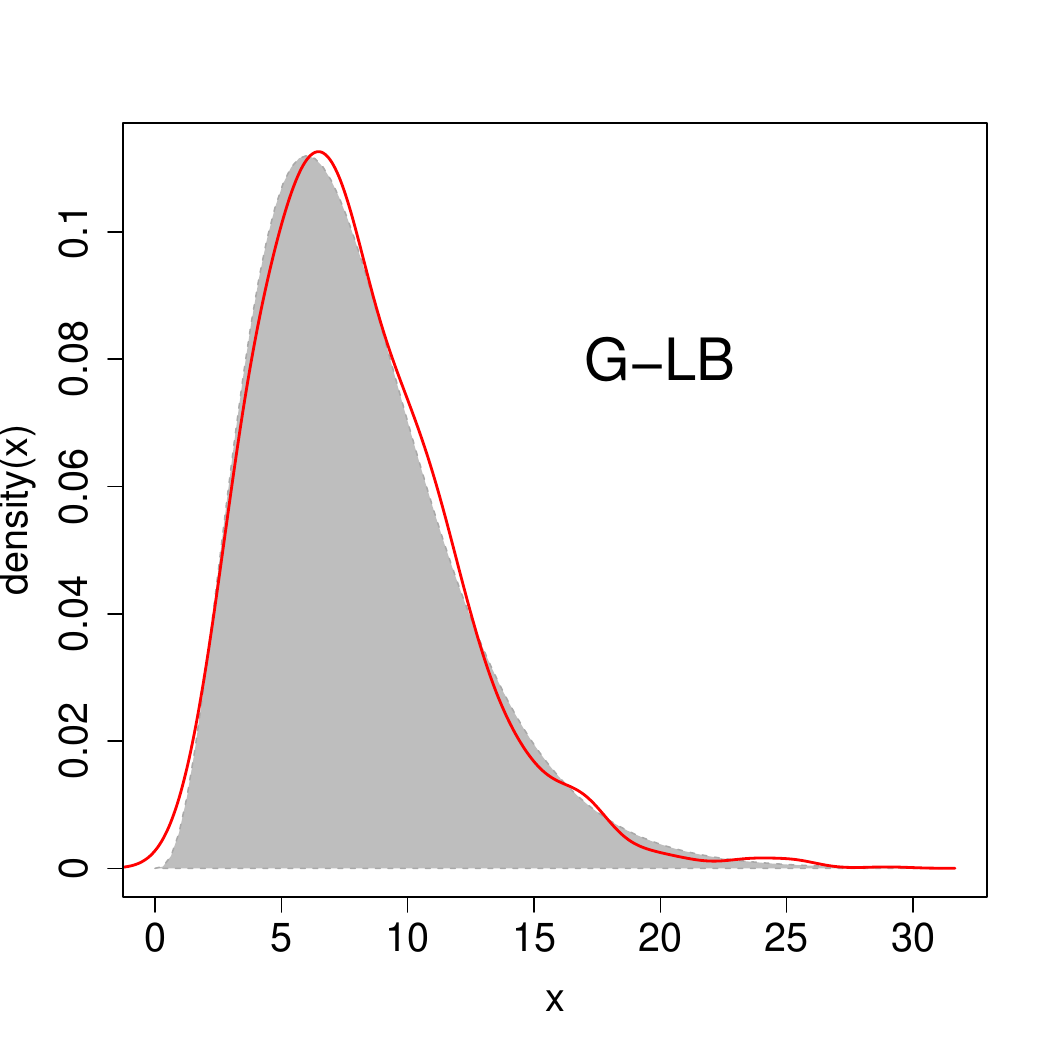}
 		\includegraphics[width = 5cm]{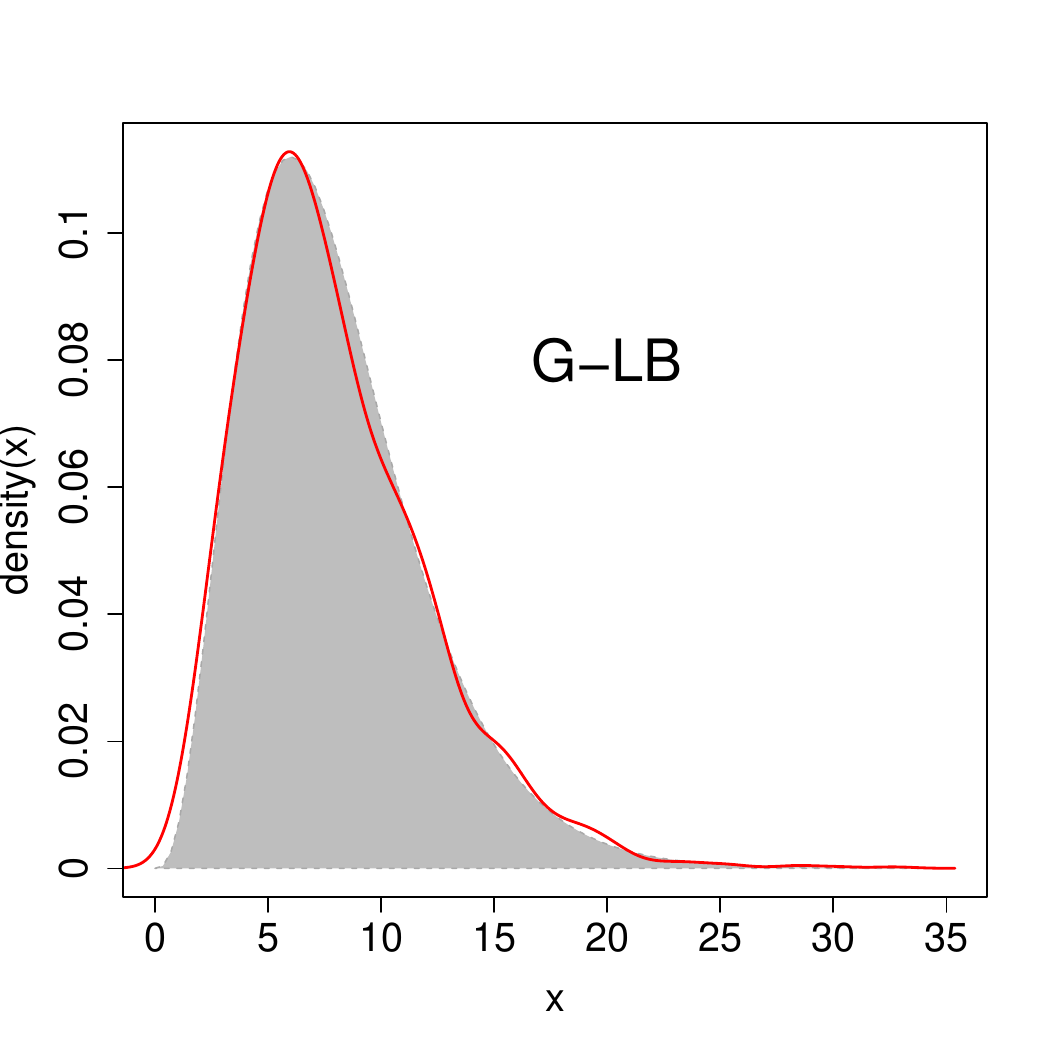}
         \includegraphics[width = 5cm]{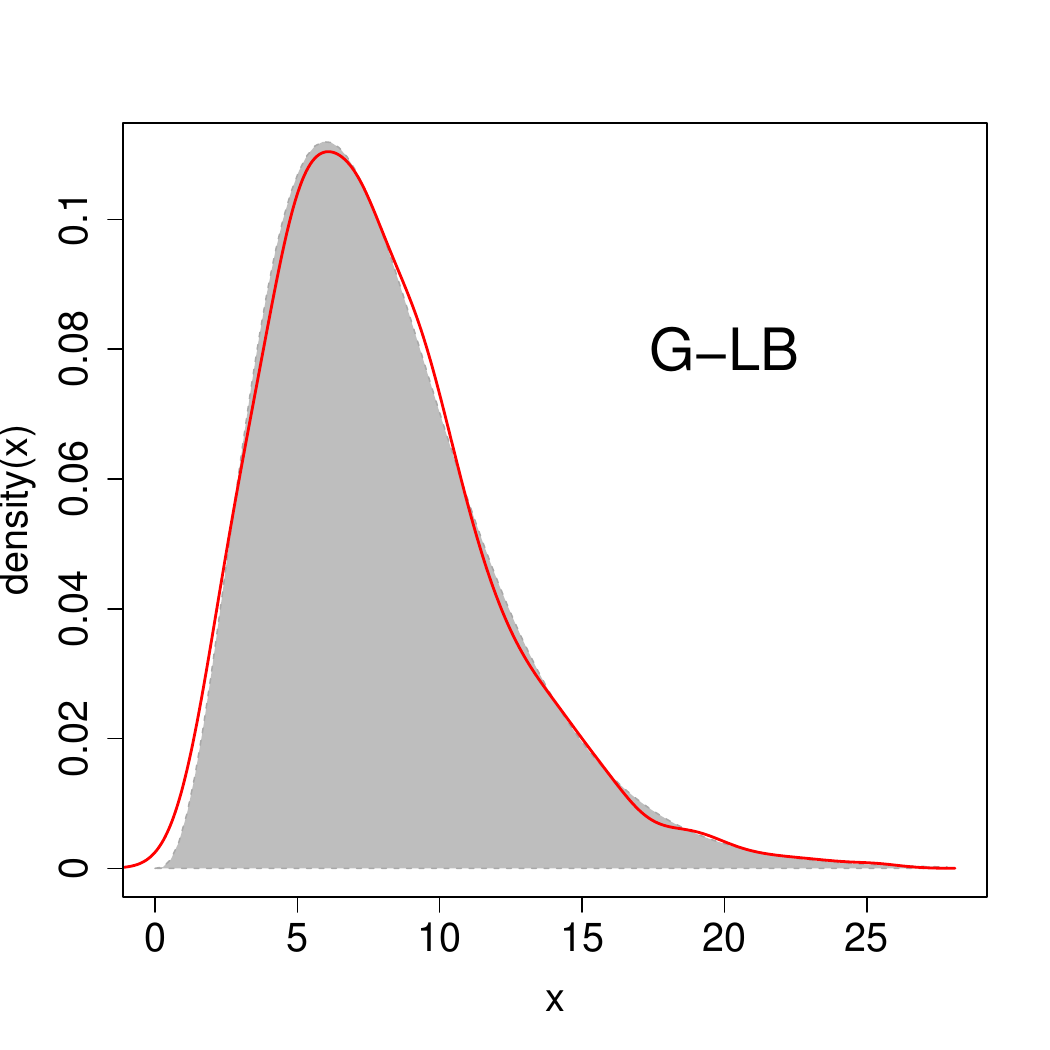}\\
 		\includegraphics[width = 5cm]{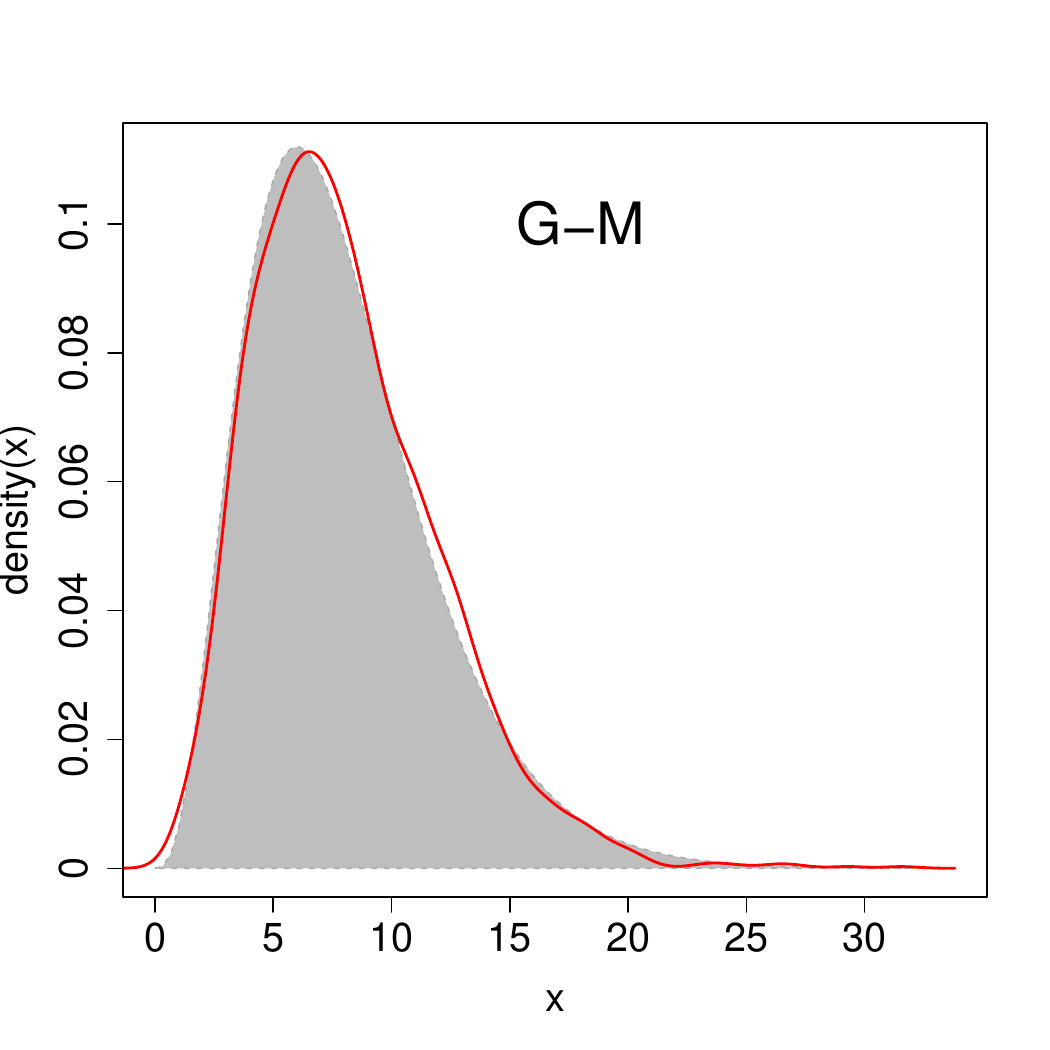}
 		\includegraphics[width = 5cm]{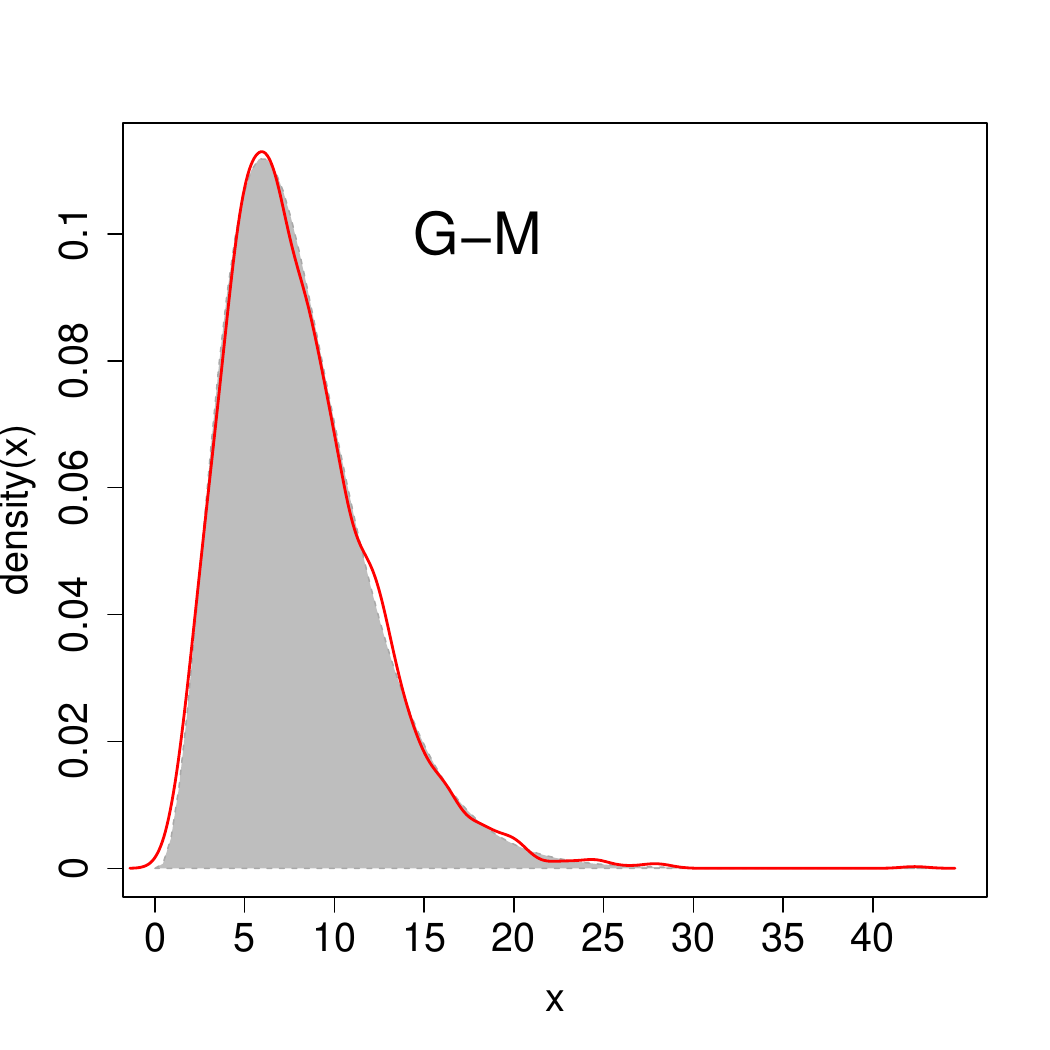}
         \includegraphics[width = 5cm]{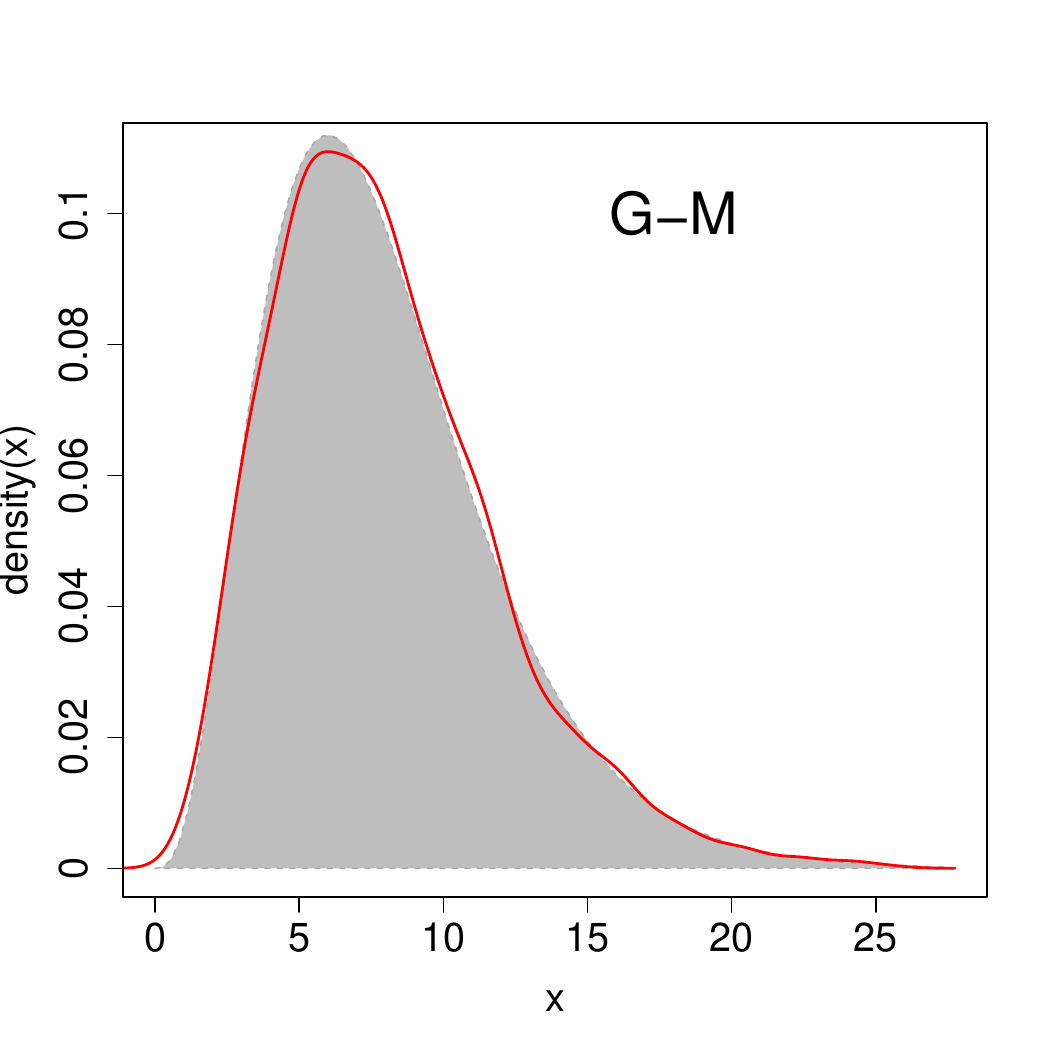}\\ 
 		\caption{Examples of empirical distributions (continuous red line) versus the reference asymptotic distributions (shaded area) for $n=250$ (first column), $n=500$ (second column) and $n=1000$ (third column). DGP is an mSARMA$(0,0)(1,0)_4(1,0)_7$ with $\Phi_{1,1}=0.5$ and $\Phi_{2,1}=0.5$.}
 		\label{test_distrib_sarsar}
 	\end{center}
\end{figure}

\begin{figure}[H]
 	\begin{center}
 		\includegraphics[width = 5cm]{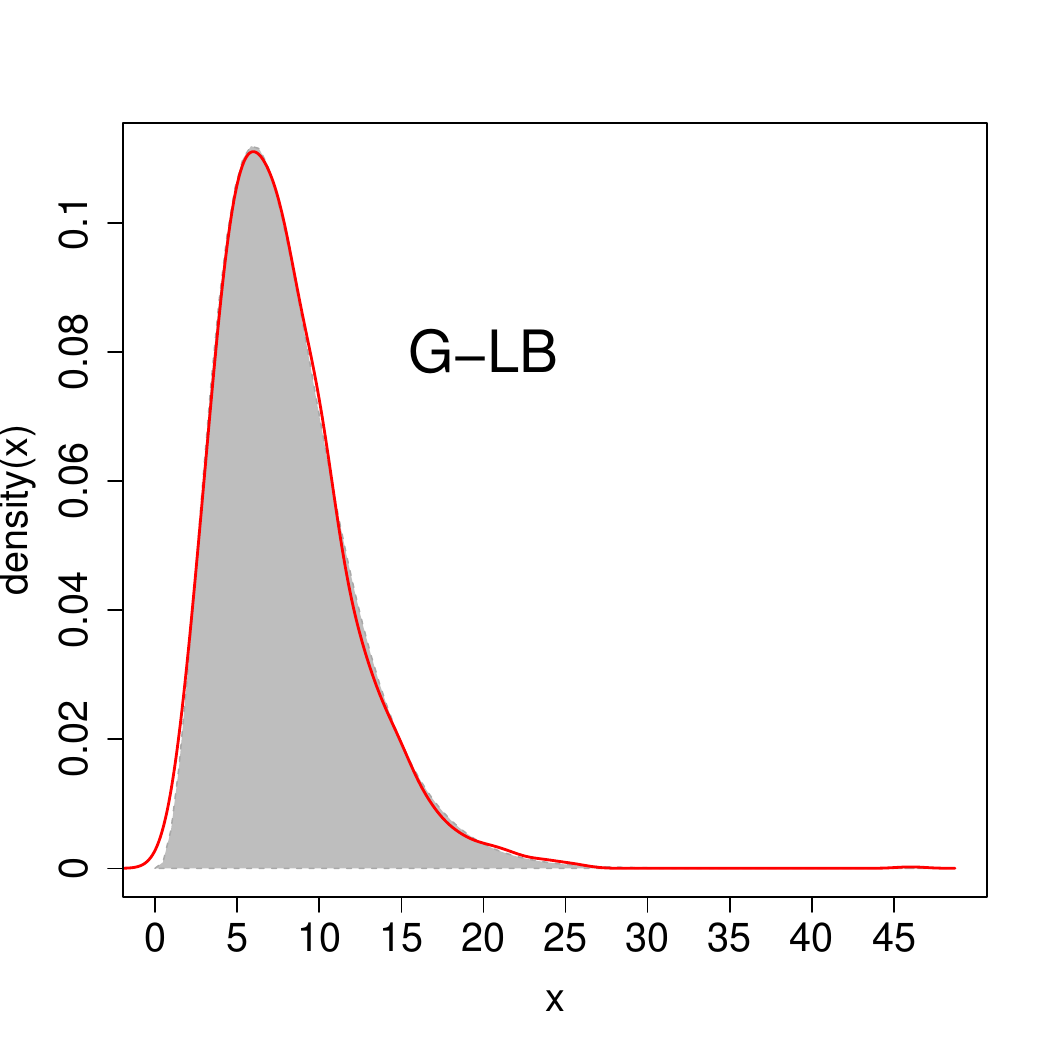}
 		\includegraphics[width = 5cm]{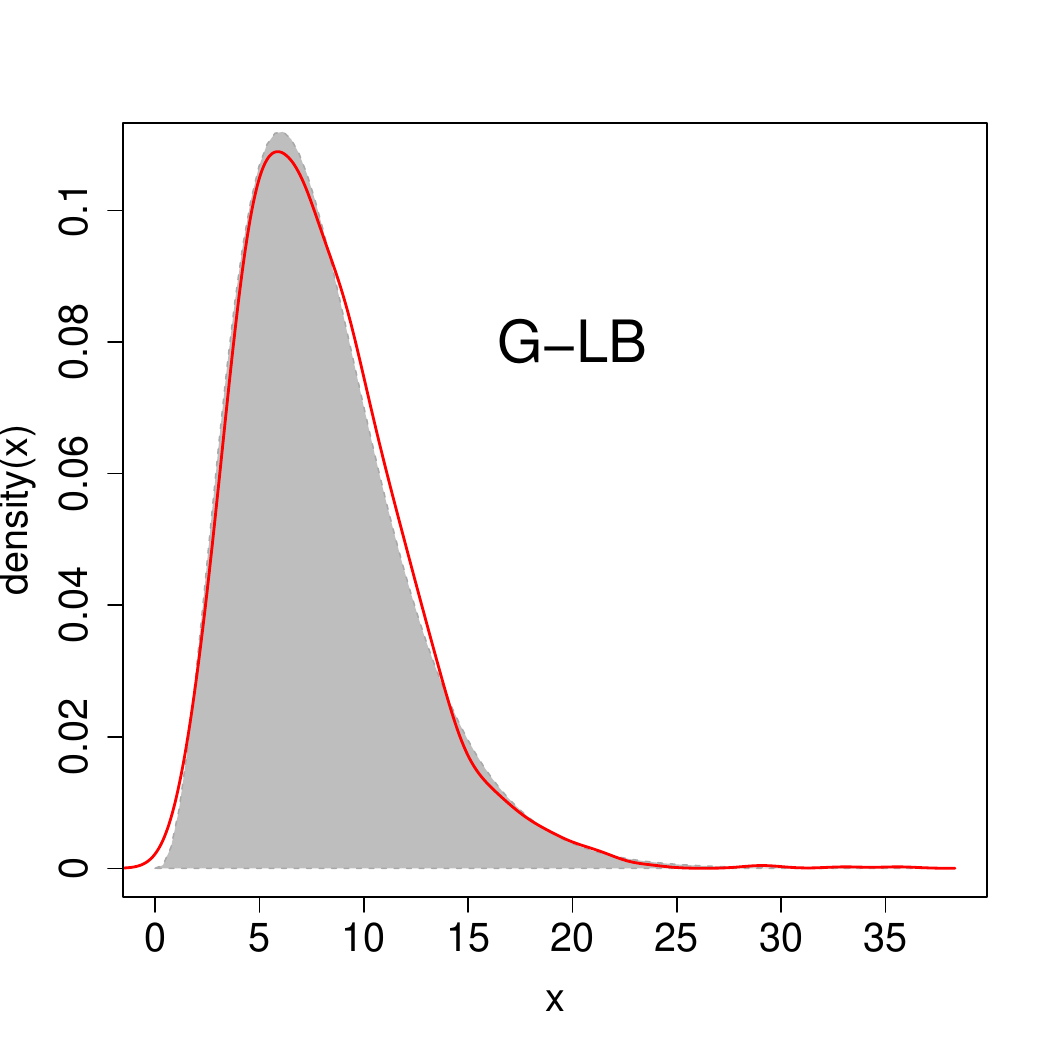}
         \includegraphics[width = 5cm]{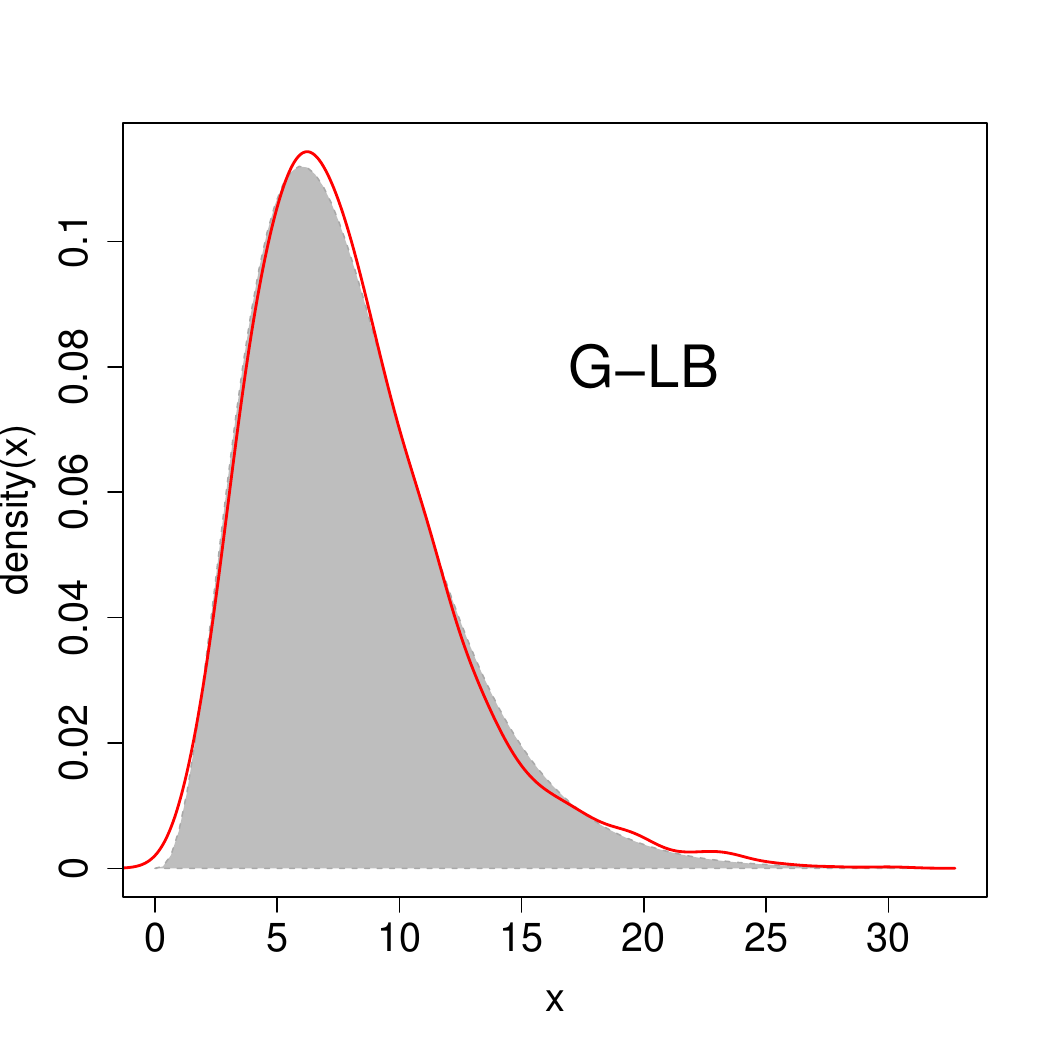}\\
 		\includegraphics[width = 5cm]{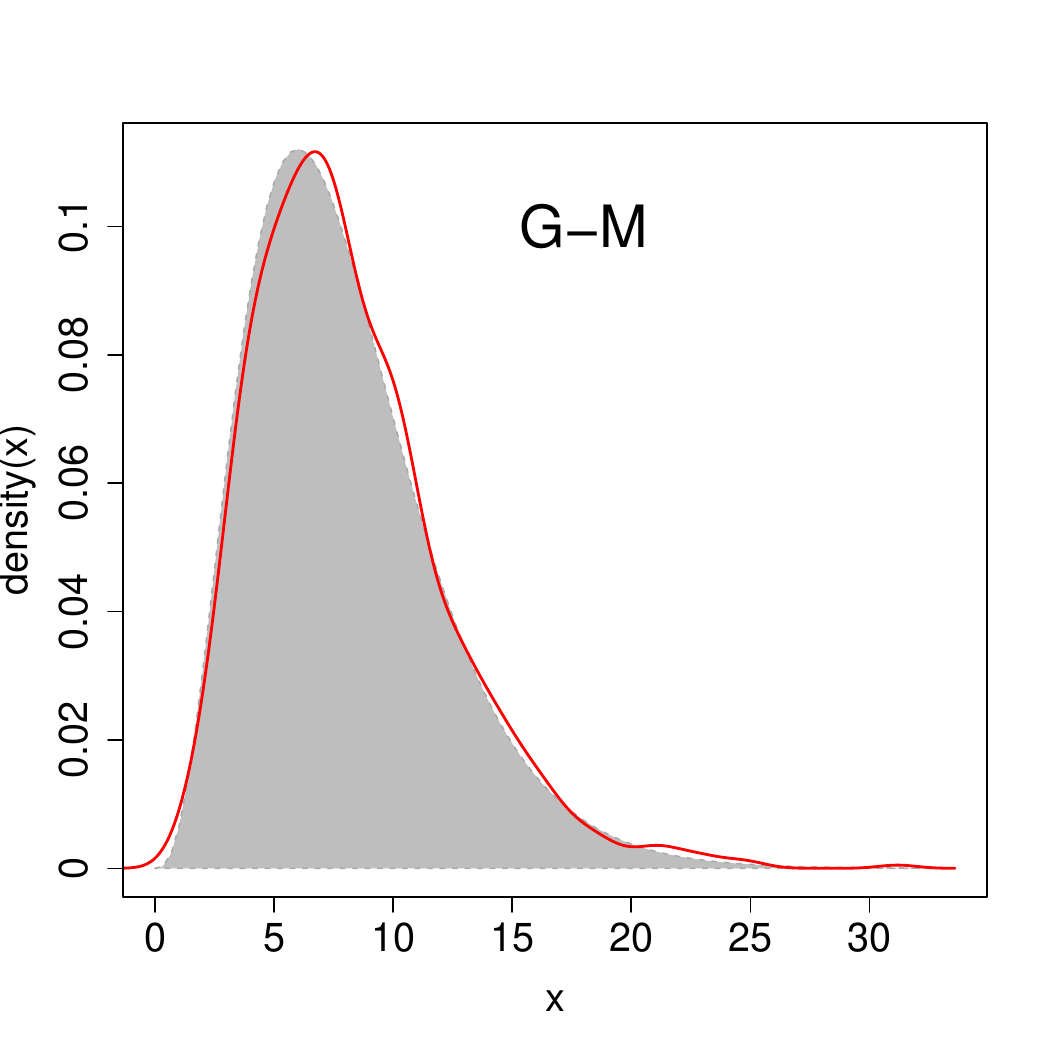}
 		\includegraphics[width = 5cm]{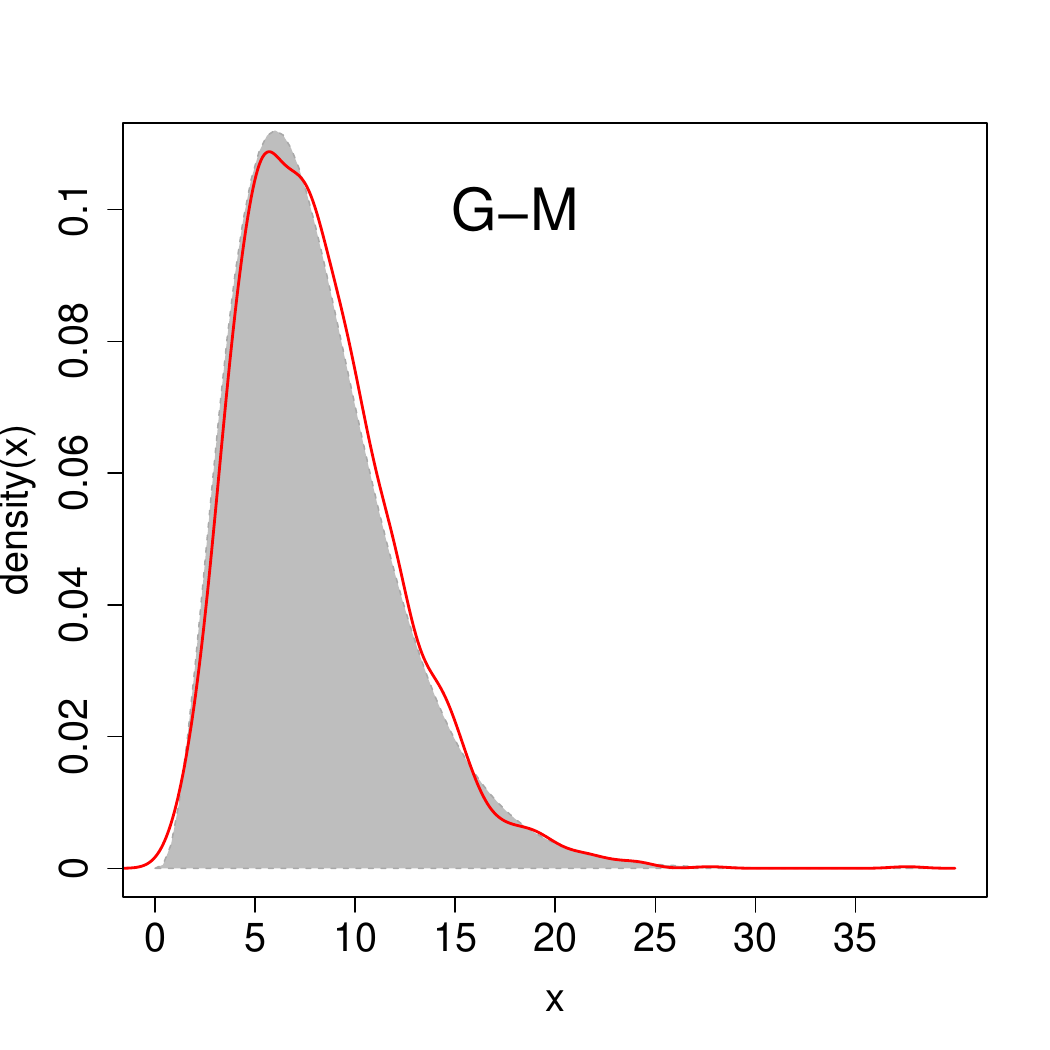}
         \includegraphics[width = 5cm]{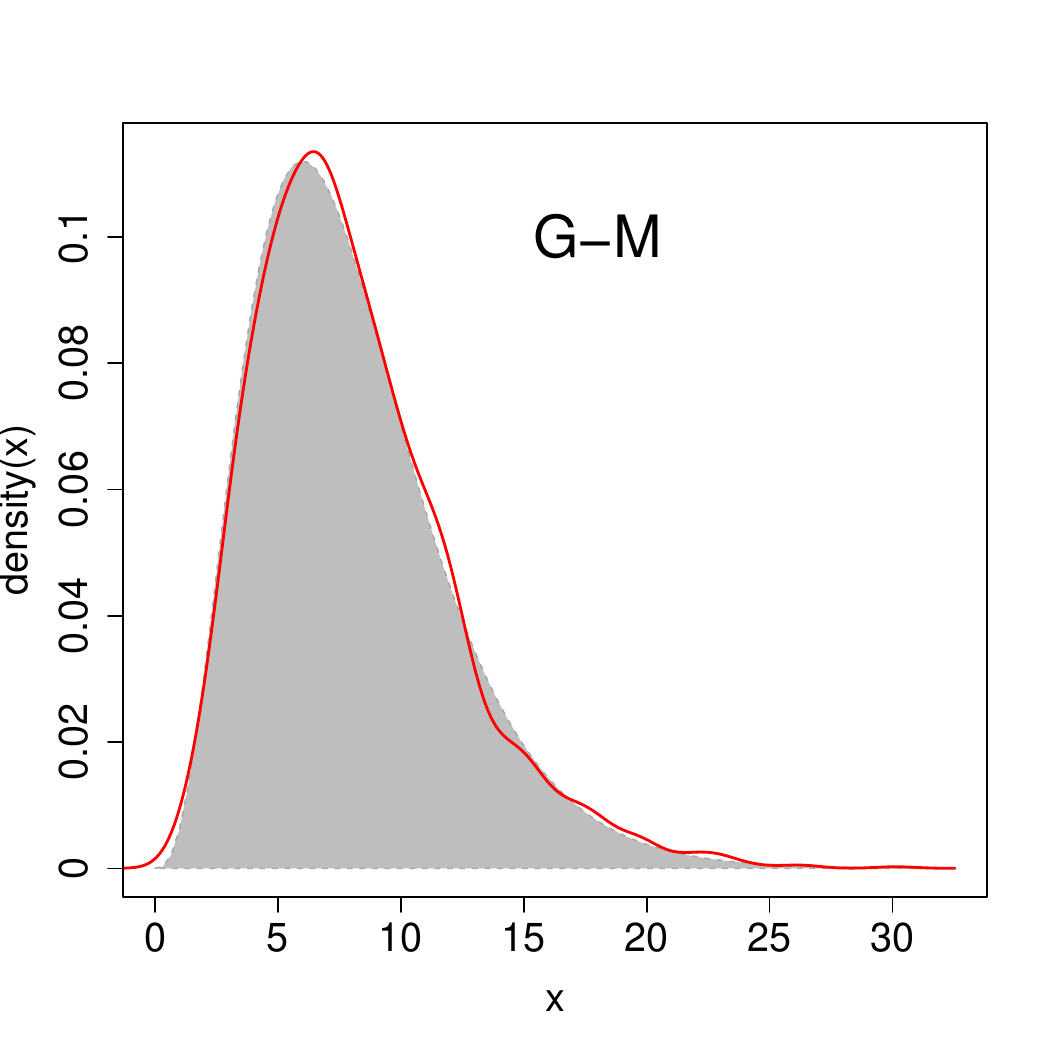}\\ 
 		\caption{Examples of empirical distributions (continuous red line) versus the reference asymptotic distributions (shaded area) for $n=250$ (first column), $n=500$ (second column) and $n=1000$ (third column). DGP is an mSARMA$(0,0)(0,1)_4(0,1)_7$ with $\Theta_{1,1}=0.5$ and $\Theta_{2,1}=0.5$.}
 		\label{test_distrib_smasma}
 	\end{center}
\end{figure}

\begin{table}[H]
	\centering
        \begin{small}
	\begin{tabular}{cc|cc|cc|cc} \hline
$\Theta_{1,1}$ & $\Theta_{2,1}$ & G-LB & G-M  & G-LB & G-M & G-LB & G-M  \\ \hline
      \multicolumn{2}{l}{$\alpha=0.01$}    & \multicolumn{2}{c}{$n=250$}  & \multicolumn{2}{c}{$n=500$} & \multicolumn{2}{c}{$n=1000$}  \\ \hline 
0.200 & 0.200 & 0.012 & 0.012 & 0.005$^{*}$ & 0.007 & 0.008 & 0.009 \\ 
  0.500 & 0.200 & 0.011 & 0.007 & 0.010 & 0.010 & 0.014 & 0.013 \\ 
  0.800 & 0.200 & 0.011 & 0.012 & 0.008 & 0.011 & 0.011 & 0.012 \\ 
  0.200 & 0.500 & 0.012 & 0.010 & 0.008 & 0.009 & 0.011 & 0.010 \\ 
  0.500 & 0.500 & 0.012 & 0.015$^{*}$ & 0.006 & 0.006 & 0.014 & 0.011 \\ 
  0.800 & 0.500 & 0.008 & 0.008 & 0.014 & 0.013 & 0.009 & 0.009 \\ 
  0.200 & 0.800 & 0.012 & 0.013 & 0.011 & 0.011 & 0.013 & 0.012 \\ 
  0.500 & 0.800 & 0.011 & 0.010 & 0.011 & 0.010 & 0.012 & 0.011 \\ 
  0.800 & 0.800 & 0.008 & 0.010 & 0.014 & 0.010 & 0.011 & 0.011 \\ 
 \hline
      \multicolumn{2}{l}{$\alpha=0.05$}    & \multicolumn{2}{c}{$n=250$}  & \multicolumn{2}{c}{$n=500$} & \multicolumn{2}{c}{$n=1000$}  \\  
 \hline
  0.200 & 0.200 & 0.052 & 0.055 & 0.044 & 0.044 & 0.052 & 0.054 \\ 
  0.500 & 0.200 & 0.047 & 0.048 & 0.046 & 0.046 & 0.059 & 0.056 \\ 
  0.800 & 0.200 & 0.044 & 0.051 & 0.044 & 0.054 & 0.044 & 0.053 \\ 
  0.200 & 0.500 & 0.040 & 0.042 & 0.046 & 0.043 & 0.050 & 0.050 \\ 
  0.500 & 0.500 & 0.046 & 0.053 & 0.039$^{*}$ & 0.050 & 0.054 & 0.051 \\ 
  0.800 & 0.500 & 0.047 & 0.052 & 0.052 & 0.052 & 0.042 & 0.043 \\ 
  0.200 & 0.800 & 0.048 & 0.054 & 0.055 & 0.057 & 0.051 & 0.052 \\ 
  0.500 & 0.800 & 0.045 & 0.052 & 0.044 & 0.046 & 0.048 & 0.048 \\ 
  0.800 & 0.800 & 0.044 & 0.047 & 0.044 & 0.054 & 0.051 & 0.054 \\ \hline
      \multicolumn{2}{l}{$\alpha=0.10$}    & \multicolumn{2}{c}{$n=250$}  & \multicolumn{2}{c}{$n=500$} & \multicolumn{2}{c}{$n=1000$}  \\ \hline 
      0.200 & 0.200 & 0.099 & 0.111 & 0.088 & 0.102 & 0.097 & 0.104 \\ 
0.500 & 0.200 & 0.090 & 0.103 & 0.093 & 0.102 & 0.104 & 0.110 \\ 
  0.800 & 0.200 & 0.090 & 0.102 & 0.094 & 0.104 & 0.102 & 0.106 \\ 
  0.200 & 0.500 & 0.088 & 0.096 & 0.090 & 0.092 & 0.099 & 0.104 \\ 
  0.500 & 0.500 & 0.094 & 0.108 & 0.082$^{**}$ & 0.092 & 0.094 & 0.094 \\ 
  0.800 & 0.500 & 0.084$^{*}$ & 0.106 & 0.100 & 0.106 & 0.092 & 0.092 \\ 
  0.200 & 0.800 & 0.101 & 0.106 & 0.099 & 0.099 & 0.098 & 0.102 \\ 
  0.500 & 0.800 & 0.094 & 0.100 & 0.101 & 0.100 & 0.102 & 0.104 \\ 
  0.800 & 0.800 & 0.087 & 0.096 & 0.098 & 0.106 & 0.093 & 0.100 \\ 
   \hline
	\end{tabular}
	\caption{Multi-seasonal case: empirical level ($\alpha_{obs}$)  of the test. Data generated from mSARIMA$(0,0,0)(0,0,1)_4 (0,0,1)_7$ model for different values of parameters $\Theta_{1,1}$ and $\Theta_{2,1}$ and sample sizes. G-LB=generalized Ljung-Box test, G-M=generalized Monti test, $\alpha$=nominal significance level, $N=2000$, $\bL=(4, 7, 8, 12, 14, 16, 20, 21, 28, 35)$, $(S_1,S_2)=(4,7)$.
	$^*$  and $^{**}$ denote empirical levels significantly different from the nominal one at $5\%$ and at $1\%$, respectively.} \label{multis_level_smasma}
    \end{small}
\end{table}
To gain further insight on the appropriateness of the asymptotic distributions and on these results, we used the 2000 values of the tests, computed on the simulated time series, to evaluate the agreement between the empirical distribution and the reference one.\\
\textcolor{black}{Figures~\ref{test_distrib_sarsar} and~\ref{test_distrib_smasma}
present representative examples of the estimated distributions for a
double SAR(1) and double SMA(1) DGP, respectively.}\\
Figures clearly show that the empirical distributions of the G-LB and
G-M test statistics nearly follow the hypothesized asymptotic
distribution.
%

\textcolor{black}{Figures~\ref{test_distrib_sarsar} and~\ref{test_distrib_smasma} provide only a few representative examples}. To assess the equivalence between the empirical and the expected distribution, for all models and for the different sample sizes we applied the Kolmogorov-Smirnov and Anderson-Darling tests of goodness-of-fit. The $p$-values of the test are reported in Table~\ref{multis_goodness_sarsar} for data generated by a double SAR(1) and in Table~\ref{multis_goodness_smasma} for a double SMA(1). The hypothesis of concordance between empirical and expected distribution is generally accepted.\\

\begin{table}[H]
	\centering
    \begin{small}
    	\begin{tabular}{cc|cc|cc|cc}
		\hline
$\Phi_{1,1}$ & $\Phi_{2,1}$ & G-LB & G-M & G-LB & G-M & G-LB & G-M  \\ \hline
      \multicolumn{2}{l}{K-S test}    & \multicolumn{2}{c}{$n=250$}  & \multicolumn{2}{c}{$n=500$} & \multicolumn{2}{c}{$n=1000$}  \\ \hline 
0.200 & 0.200 & 0.329 & 0.001$^{**}$ & 0.611 & 0.068 & 0.012$^{*}$ & 0.267 \\ 
  0.500 & 0.200 & 0.228 & 0.323 & 0.349 & 0.601 & 0.966 & 0.892 \\ 
  0.800 & 0.200 & 0.607 & 0.015$^{*}$ & 0.853 & 0.472 & 0.117 & 0.083 \\ 
  0.200 & 0.500 & 0.606 & 0.409 & 0.765 & 0.116 & 0.502 & 0.784 \\ 
  0.500 & 0.500 & 0.543 & 0.165 & 0.082 & 0.792 & 0.829 & 0.567 \\ 
  0.800 & 0.500 & 0.109 & 0.102 & 0.470 & 0.131 & 0.086 & 0.042$^{*}$ \\ 
  0.200 & 0.800 & 0.193 & 0.003$^{**}$ & 0.871 & 0.218 & 0.322 & 0.343 \\ 
  0.500 & 0.800 & 0.176 & 0.412 & 0.504 & 0.733 & 0.412 & 0.159 \\ 
0.800 & 0.800 & 0.376 & 0.055 & 0.133 & 0.013$^{*}$ & 0.436 & 0.273 \\ \hline 
      \multicolumn{2}{l}{A-D test}    & \multicolumn{2}{c}{$n=250$}  & \multicolumn{2}{c}{$n=500$} & \multicolumn{2}{c}{$n=1000$}  \\ \hline 
0.200 & 0.200 & 0.463 & 0.008$^{**}$ & 0.595 & 0.060 & 0.021 & 0.144 \\ 
  0.500 & 0.200 & 0.124 & 0.212 & 0.430 & 0.484 & 0.942 & 0.944 \\ 
  0.800 & 0.200 & 0.806 & 0.024$^{*}$ & 0.762 & 0.645 & 0.046$^{*}$ & 0.025$^{*}$ \\ 
  0.200 & 0.500 & 0.559 & 0.151 & 0.421 & 0.070 & 0.618 & 0.861 \\ 
  0.500 & 0.500 & 0.419 & 0.101 & 0.139 & 0.974 & 0.650 & 0.458 \\ 
  0.800 & 0.500 & 0.179 & 0.112 & 0.610 & 0.107 & 0.116 & 0.039$^{*}$ \\ 
  0.200 & 0.800 & 0.214 & 0.018$^{*}$ & 0.797 & 0.190 & 0.179 & 0.106 \\ 
  0.500 & 0.800 & 0.197 & 0.448 & 0.552 & 0.864 & 0.363 & 0.068 \\ 
  0.800 & 0.800 & 0.365 & 0.018$^{*}$ & 0.075 & 0.004$^{**}$ & 0.321 & 0.173 \\ 
   \hline
	\end{tabular}
	\caption{$P$-values of the Kolmogorov-Smirnov and Anderson-Darling goodness-of-fit tests for mSARIMA$(0,0,0)(1,0,0)_4 (1,0,0)_7$ DGP with different values of parameters $\Phi_{1,1}$ and $\Phi_{2,1}$ and sample sizes. G-LB=generalized Ljung-Box test, G-M=generalized Monti test. $\alpha$=nominal significance level, $N=2000$.     $^*$  and $^{**}$ denote rejection of the hypothesis of correct distribution at $5\%$ and at $1\%$, respectively.} \label{multis_goodness_sarsar}
    \end{small}
\end{table}
\begin{table}[H]
	\centering
    \begin{small}
	\begin{tabular}{cc|cc|cc|cc}
		\hline
$\Theta_{1,1}$ & $\Theta_{2,1}$ & G-LB & G-M & G-LB & G-M & G-LB & G-M  \\ \hline
      \multicolumn{2}{l}{K-S test}    & \multicolumn{2}{c}{$n=250$}  & \multicolumn{2}{c}{$n=500$} & \multicolumn{2}{c}{$n=1000$}  \\ \hline 
0.200 & 0.200 & 0.631 & 0.017$^{*}$ & 0.586 & 0.804 & 0.480 & 0.445 \\ 
  0.500 & 0.200 & 0.018$^{*}$ & 0.481 & 0.411 & 0.061 & 0.882 & 0.801 \\ 
  0.800 & 0.200 & 0.361 & 0.003$^{**}$ & 0.620 & 0.194 & 0.360 & 0.081 \\ 
  0.200 & 0.500 & 0.408 & 0.101 & 0.568 & 0.205 & 0.911 & 0.449 \\ 
  0.500 & 0.500 & 0.458 & 0.122 & 0.237 & 0.926 & 0.663 & 0.696 \\ 
  0.800 & 0.500 & 0.103 & 0.100 & 0.819 & 0.231 & 0.893 & 0.433 \\ 
  0.200 & 0.800 & 0.373 & 0.098 & 0.201 & 0.407 & 0.629 & 0.993 \\ 
  0.500 & 0.800 & 0.068 & 0.025 & 0.655 & 0.128 & 0.902 & 0.805 \\ 
0.800 & 0.800 & 0.015 $^{*}$ & 0.101 & 0.916 & 0.012$^{*}$ & 0.519 & 0.275 \\  \hline
      \multicolumn{2}{l}{A-D test}    & \multicolumn{2}{c}{$n=250$}  & \multicolumn{2}{c}{$n=500$} & \multicolumn{2}{c}{$n=1000$}  \\ \hline 
0.200 & 0.200 & 0.693 & 0.032$^{*}$ & 0.218 & 0.657 & 0.373 & 0.409 \\ 
  0.500 & 0.200 & 0.008$^{*}$ & 0.707 & 0.246 & 0.021$^{*}$ & 0.725 & 0.664 \\ 
  0.800 & 0.200 & 0.130 & 0.001$^{**}$ & 0.581 & 0.440 & 0.408 & 0.052 \\ 
  0.200 & 0.500 & 0.406 & 0.106 & 0.369 & 0.096 & 0.955 & 0.657 \\ 
  0.500 & 0.500 & 0.489 & 0.056 & 0.113 & 0.968 & 0.278 & 0.446 \\ 
  0.800 & 0.500 & 0.037$^{*}$ & 0.214 & 0.939 & 0.254 & 0.642 & 0.497 \\ 
  0.200 & 0.800 & 0.251 & 0.058 & 0.353 & 0.536 & 0.685 & 0.977 \\ 
  0.500 & 0.800 & 0.081 & 0.016 & 0.583 & 0.088 & 0.795 & 0.647 \\ 
  0.800 & 0.800 & 0.002$^{**}$ & 0.152 & 0.849 & 0.001$^{**}$ & 0.391 & 0.118 \\ 
   \hline
	\end{tabular}
	\caption{$P$-values of the Kolmogorov-Smirnov and Anderson-Darling goodness-of-fit tests for mSARIMA$(0,0,0)(0,0,1)_4 (0,0,1)_7$ DGP for different values of parameters $\Theta_{1,1}$ and $\Theta_{2,1}$ and sample sizes. G-LB=generalized Ljung-Box test, G-M=generalized Monti test. $N=2000$.     $^*$  and $^{**}$ denote rejection of the hypothesis of correct distribution at $5\%$ and at $1\%$, respectively.} \label{multis_goodness_smasma}
    \end{small}
\end{table}
We conclude the analyses of this section by investigating the power of the tests. Table~\ref{multiciclo_power} lists the rejection rate for non-correctly specified models.
The data-generating processes are specified by the parameter values reported
in the first six columns. The tests are then applied to the residuals
of a fitted (simple) SAR(1) model and of a (simple) SMA(1) model.\\
Whenever the weight of the neglected components is not too small, the power of the tests is generally very high. On the whole, these power levels are coherent with those reported in the literature for the original and the seasonal versions of the tests and, again, suggest the suitability of the generalized versions. 
	\begin{table}[H]
    \centering
	\begin{small}
\begin{tabular}
{p{0.8cm}p{0.8cm}p{0.8cm}p{0.8cm}p{0.8cm}p{0.8cm}|p{0.9cm}p{0.9cm}|p{0.9cm}p{0.9cm}|p{0.9cm}p{0.9cm}} 
\hline
    $\phi_1$  &  $\theta_1$ & $\Phi_{1,1}$ & $\Phi_{2,1}$ & $\Theta_{1,1}$ & $\Theta_{2,1}$ & G-LB & G-M  &  G-LB & G-M & G-LB & G-M   \\ \hline  
 \multicolumn{6}{l}{Fitted model: mSARMA(0,0)(1,0)$_4$(0,0)$_7$}  & \multicolumn{2}{c}{$n=250$}    & \multicolumn{2}{c}{$n=500$}  & \multicolumn{2}{c}{$n=1000$}  \\ \hline 
 * & * & 0.5 & 0.5 & *  & * & 1.000 & 1.000 & 1.000 & 1.000 & 1.000 & 1.000 \\ 
  * & *  & 0.5 & 0.3 & *  & * & 0.921 & 0.908 & 0.999 & 0.999 & 1.000 & 1.000 \\ 
  * & *  & 0.5 & 0.1 & *  & * & 0.139 & 0.117 & 0.274 & 0.262 & 0.541 & 0.528 \\ 
  * & *  & 0.5 & *  & * & 0.5 & 1.000 & 1.000 & 1.000 & 1.000 & 1.000 & 1.000 \\ 
  * & *  & 0.5 & *  & * & 0.3 & 0.881 & 0.888 & 1.000 & 1.000 & 1.000 & 1.000 \\ 
  * & *  & 0.5 & *  & * & 0.1 & 0.129 & 0.122 & 0.268 & 0.265 & 0.534 & 0.544 \\ 
  * & *  & 0.5 & *  & 0.5 & *  & 0.987 & 0.992 & 1.000 & 1.000 & 1.000 & 1.000 \\ 
  * & *  & 0.5 & *  & 0.3 & *  & 0.540 & 0.534 & 0.899 & 0.904 & 0.999 & 0.999 \\ 
   \hline
   \multicolumn{6}{l}{Fitted model:  mSARMA(0,0)(1,0)$_4$(0,0)$_{12}$}  & \multicolumn{2}{c}{$n=250$}    & \multicolumn{2}{c}{$n=500$}  & \multicolumn{2}{c}{$n=1000$}  \\ \hline 
    * & * & 0.5 & 0.5 & * & * & 0.901 & 0.944 & 1.000 & 1.000 & 1.000 & 1.000 \\ 
  * & * & 0.5 & 0.3 & * & * & 0.324 & 0.359 & 0.999 & 0.998 & 1.000 & 1.000 \\ 
  * & * & 0.5 & 0.1 & * & * & 0.141 & 0.153 & 0.255 & 0.239 & 0.556 & 0.533 \\ 
  * & * & 0.5 & * & * & 0.5 & 1.000 & 1.000 & 1.000 & 1.000 & 1.000 & 1.000 \\ 
  * & * & 0.5 & * & * & 0.3 & 1.000 & 1.000 & 0.998 & 0.998 & 1.000 & 1.000 \\ 
  * & * & 0.5 & * & * & 0.1 & 0.149 & 0.161 & 0.248 & 0.236 & 0.530 & 0.521 \\ 
  * & * & 0.5 & * & 0.5 & * & 0.997 & 0.999 & 1.000 & 1.000 & 1.000 & 1.000 \\ 
  * & * & 0.5 & * & 0.3 & * & 0.819 & 0.807 & 0.909 & 0.917 & 0.999 & 0.999 \\  \hline
   \multicolumn{6}{l}{Fitted model:  mSARMA(0,0)(0,1)$_4$(0,0)$_7$ }   & \multicolumn{2}{c}{$n=250$}    & \multicolumn{2}{c}{$n=500$}  & \multicolumn{2}{c}{$n=1000$}  \\ \hline 
* & *  & * & *  & 0.5 & 0.5 & 1.000 & 1.000 & 1.000 & 1.000 & 1.000 & 1.000 \\ 
  * & *  & * & *  & 0.5 & 0.3 & 0.883 & 0.897 & 0.998 & 0.999 & 1.000 & 1.000 \\ 
  * & *  & * & *  & 0.5 & 0.1 & 0.126 & 0.122 & 0.251 & 0.250 & 0.536 & 0.542 \\ 
  * & *  & * & 0.5 & 0.5 & * & 1.000 & 1.000 & 1.000 & 1.000 & 1.000 & 1.000 \\ 
  * & *  & * & 0.3 & 0.5 & *  & 0.917 & 0.905 & 1.000 & 0.999 & 1.000 & 1.000 \\ 
  * & *  & * & 0.1 & 0.5 & *  & 0.130 & 0.130 & 0.268 & 0.261 & 0.549 & 0.546 \\ 
  * & *  & 0.5 & *  & 0.5 & *  & 0.996 & 0.994 & 1.000 & 1.000 & 1.000 & 1.000 \\ 
  * & *  & 0.3 & *  & 0.5 & *  & 0.563 & 0.526 & 0.908 & 0.899 & 1.000 & 1.000 \\    
   \hline
  \multicolumn{6}{l}{Fitted model: mSARMA(0,0)(0,1)$_4$(0,0)$_{12}$}  & \multicolumn{2}{c}{$n=250$}    & \multicolumn{2}{c}{$n=500$}  & \multicolumn{2}{c}{$n=1000$}  \\ \hline 
  * & * & * & * & 0.5 & 0.5 & 1.000 & 1.000 & 1.000 & 1.000 & 1.000 & 1.000 \\ 
  * & * & * & * & 0.5 & 0.3 & 0.848 & 0.852 & 0.999 & 0.999 & 1.000 & 1.000 \\
  * & * & * & * & 0.5 & 0.1 & 0.134 & 0.106 & 0.247 & 0.232 & 0.527 & 0.517 \\ 
  * & * & * & 0.5 & 0.5 & * & 1.000 & 1.000 & 1.000 & 1.000 & 1.000 & 1.000 \\ 
  * & * & * & 0.3 & 0.5 & * & 0.904 & 0.863 & 1.000 & 1.000 & 1.000 & 1.000 \\ 
  * & * & * & 0.1 & 0.5 & * & 0.148 & 0.116 & 0.267 & 0.241 & 0.544 & 0.524 \\ 
  * & * & 0.5 & * & 0.5 & * & 0.996 & 0.994 & 1.000 & 1.000 & 1.000 & 1.000 \\ 
  * & * & 0.3 & * & 0.5 & * & 0.585 & 0.541 & 0.917 & 0.904 & 1.000 & 1.000 \\  \hline
   \end{tabular}
\caption{Empirical power for multi-seasonal portmanteau tests. The first six columns define the DGP. The fitted models are a simple SAR(1) and a simple SMA(1) for $(m_1,m_2)=(4,7)$ and $(4,12)$.} \label{multiciclo_power}
\end{small}
\end{table}



\subsection{Portmanteau tests for mixed (seasonal and non-seasonal) residual autocorrelation}

In this section, we explore the performance of the generalized portmanteau approach for testing the system of hypotheses~(\ref{sys_hyp}), but with vector $\bL$ jointly considering short-term and seasonal residual autocorrelation.\\
When investigating the type-I error, we generated data from  mSARIMA models with two seasonal components, of periods $(S_1, S_2)=(4,7)$  and $(4,12)$ and a short-memory component. In this case, we fit the correct model and we apply the tests using always $m_0=m_1=m_2=5$, i.e.\ we consider five lags for each component. Clearly, these lags can overlap.  
Table~\ref{mixed_level_47} lists the empirical test levels versus the nominal ones for ten different mSARIMA models and highlights with asterisks the cases for which the observed level is significantly different from the nominal one. \\
Overall, results emerging from Table~\ref{mixed_level_47} are similar to those obtained in Section~\ref{sect:multiseas}.
For the generalized version of the Ljung-Box and Monti tests the
observed levels are coherent with the nominal level.
%
\begin{table}[H]
\centering
\begin{small}
   	\begin{tabular}{p{0.6cm}p{0.6cm}p{0.6cm}p{0.6cm}p{0.6cm}p{0.6cm}|p{1.0cm}p{1.0cm}|p{1.0cm}p{1.0cm}|p{1.0cm}p{1.0cm}} \hline
$\phi_{1}$ & $\theta_{1}$ & $\Phi_{1,1}$ & $\Phi_{2,1}$ & $\Theta_{1,1}$ & $\Theta_{2,1}$ & G-LB & G-M &  G-LB & G-M &  G-LB & G-M  \\ \hline
      \multicolumn{6}{l}{$\alpha=0.01$}    & \multicolumn{2}{c}{$n=250$}  & \multicolumn{2}{c}{$n=500$} & \multicolumn{2}{c}{$n=1000$}  \\ \hline 
  0.5& * & 0.5& 0.3& * & * & 0.011 & 0.012 & 0.009 & 0.006 & 0.009 & 0.007 \\ 
  0.1& * & 0.5& 0.3& * & * & 0.010 & 0.008 & 0.011 & 0.011 & 0.010 & 0.011 \\ 
  * & 0.5& 0.5& 0.3& * & * & 0.014 & 0.012 & 0.010 & 0.010 & 0.012 & 0.012 \\ 
  * & 0.1& 0.5& 0.3& * & * & 0.009 & 0.012 & 0.014 & 0.013 & 0.011 & 0.009 \\ 
  0.5& * & * & * & 0.5& 0.3& 0.009 & 0.010 & 0.009 & 0.010 & 0.007 & 0.008 \\ 
  0.1& * & * & * & 0.5& 0.3& 0.018$^{**}$ & 0.018$^{**}$ & 0.013 & 0.012 & 0.010 & 0.008 \\ 
  * & 0.5& * & * & 0.5& 0.3& 0.007 & 0.010 & 0.011 & 0.010 & 0.015 & 0.012 \\ 
  * & 0.1& * & * & 0.5& 0.3& 0.008 & 0.011 & 0.007 & 0.009 & 0.010 & 0.011 \\ 
  0.3& * & 0.5& * & * & 0.4& 0.007 & 0.011 & 0.007 & 0.009 & 0.012 & 0.011 \\ 
  * & 0.3& 0.5& * & * & 0.4& 0.011 & 0.010 & 0.011 & 0.008 & 0.006 & 0.004$^{**}$ \\ \hline
      \multicolumn{6}{l}{$\alpha=0.05$}    & \multicolumn{2}{c}{$n=250$}  & \multicolumn{2}{c}{$n=500$} & \multicolumn{2}{c}{$n=1000$}  \\ \hline 
0.5& * & 0.5& 0.3& * & * & 0.050 & 0.056 & 0.048 & 0.050 & 0.051 & 0.046 \\ 
  0.1& * & 0.5& 0.3& * & * & 0.043 & 0.047 & 0.048 & 0.050 & 0.052 & 0.050 \\ 
  * & 0.5& 0.5& 0.3& * & * & 0.059 & 0.059 & 0.058 & 0.055 & 0.046 & 0.052 \\ 
  * & 0.1& 0.5& 0.3& * & * & 0.058 & 0.054 & 0.052 & 0.054 & 0.051 & 0.049 \\ 
  0.5& * & * & * & 0.5& 0.3& 0.043 & 0.048 & 0.052 & 0.051 & 0.052 & 0.051 \\ 
  0.1& * & * & * & 0.5& 0.3& 0.058 & 0.066$^{**}$ & 0.056 & 0.057 & 0.052 & 0.052 \\ 
  * & 0.5& * & * & 0.5& 0.3& 0.044 & 0.044 & 0.043 & 0.052 & 0.053 & 0.054 \\ 
  * & 0.1& * & * & 0.5& 0.3& 0.054 & 0.051 & 0.048 & 0.052 & 0.054 & 0.054 \\ 
  0.3& * & 0.5& * & * & 0.4& 0.054 & 0.049 & 0.048 & 0.047 & 0.052 & 0.051 \\
  * & 0.3& 0.5& * & * & 0.4& 0.052 & 0.052 & 0.048 & 0.047 & 0.045 & 0.047 \\ \hline
      \multicolumn{6}{l}{$\alpha=0.10$}    & \multicolumn{2}{c}{$n=250$}  & \multicolumn{2}{c}{$n=500$} & \multicolumn{2}{c}{$n=1000$}  \\ \hline 
  0.5& * & 0.5& 0.3& * & * & 0.100& 0.108 & 0.097 & 0.102 & 0.090 & 0.096 \\ 
  0.1& * & 0.5& 0.3& * & * & 0.090 & 0.098 & 0.097 & 0.102 & 0.105 & 0.106 \\ 
  * & 0.5& 0.5& 0.3& * & * & 0.104 & 0.112 & 0.108 & 0.108 & 0.102 & 0.102 \\ 
  * & 0.1& 0.5& 0.3& * & * & 0.110 & 0.108 & 0.110 & 0.119$^{**}$ & 0.101 & 0.099 \\ 
  0.5& * & * & * & 0.5& 0.3& 0.100& 0.100& 0.094 & 0.100& 0.107 & 0.104 \\ 
  0.1& * & * & * & 0.5& 0.3& 0.108 & 0.121$^{**}$ & 0.104 & 0.110 & 0.097 & 0.100\\ 
  * & 0.5& * & * & 0.5& 0.3& 0.092 & 0.092 & 0.094 & 0.095 & 0.094 & 0.097 \\ 
  * & 0.1& * & * & 0.5& 0.3& 0.108 & 0.112 & 0.095 & 0.101 & 0.107 & 0.104 \\ 
  0.3& * & 0.5& * & * & 0.4& 0.106 & 0.110 & 0.094 & 0.102 & 0.108 & 0.104 \\ 
  * & 0.3& 0.5& * & * & 0.4& 0.102 & 0.114 & 0.088 & 0.094 & 0.096 & 0.094 \\ 
   \hline
	\end{tabular}
	\caption{Empirical test level ($\alpha_{obs}$) for different data generating processes, defined by parameters in the first six columns, sample sizes and nominal levels. G-LB=generalized Ljung-Box test, G-M=generalized Monti test. $(S_1,S_2)=(4,7)$,  $m_0=m_1=m_2=5$.
    $^*$  and $^{**}$ denote empirical levels significantly different from the nominal one at $5\%$ and at $1\%$, respectively.} \label{mixed_level_47}
    \end{small}
\end{table}
%
Results on the power of the tests, at the $5\%$ level, are summarized in Table \ref{mixed_pow_47}. Data are generated by mSARMA$(p,d)(P_1,Q_1)_{4}(P_2,Q_2)_{7}$ models for $n=250,500$ and $1000$ but the fitted models erroneously neglect the short-term components. \\
Table~\ref{mixed_pow_47} is divided into two blocks: the upper block lists the empirical power when we test the system of hypotheses~(\ref{sys_hyp}) with lag vector $\bL=(1,2,3,4,5,7,8,12,14,16,20,21,28,35)$. In this case we explicitly consider the possibility of joint non-seasonal and seasonal residual autocorrelation. 
The numbers in the Table show that all tests have extremely high power and clearly "see" the residual autocorrelation.\\
The results in the lower block of the table refer to a system of hypotheses which assumes no autocorrelation at the seasonal lags of the two periodic components, without accounting for short-term autocorrelation. This means considering $m_0=0$ and $m_1=m_2=5$ and thus $\bL=(4,7,8,12,14,16,20,21,28,35)$.
Here, the neglected short-term component is such that it does not impact the structure of the seasonal autocorrelation so that the hypothesis of no seasonal autocorrelation is basically true. In this second framework, the power is clearly lower but yet much higher than the nominal test level.

\begin{table}[H]
\centering
\begin{small}
   	\begin{tabular}{p{0.6cm}p{0.6cm}p{0.6cm}p{0.6cm}p{0.6cm}p{0.6cm}|p{1.0cm}p{1.0cm}|p{1.0cm}p{1.0cm}|p{1.0cm}p{1.0cm}} \hline
$\phi_{1}$ & $\theta_{1}$ & $\Phi_{1,1}$ & $\Phi_{2,1}$ & $\Theta_{1,1}$ & $\Theta_{2,1}$ & G-LB & G-M &  G-LB & G-M  & G-LB & G-M \\ \hline
\multicolumn{6}{l}{$\alpha=0.05$, \ $m_0=m_1=m_2=5$} & \multicolumn{2}{c}{$n=250$} & \multicolumn{2}{c}{$n=500$}& \multicolumn{2}{c}{$n=1000$} \\ \hline
0.5 & * & 0.5 & 0.3 & * & * & 1.000 & 1.000 & 1.000 & 1.000 & 1.000 & 1.000 \\ 
  0.3 & * & 0.5 & 0.3 & * & * & 0.922 & 0.925 & 1.000 & 1.000 & 1.000 & 1.000 \\ 
  0.1 & * & 0.5 & 0.3 & * & * & 0.170 & 0.176 & 0.295 & 0.294 & 0.554 & 0.556 \\ 
  * & 0.5 & 0.5 & 0.3 & * & * & 1.000 & 1.000 & 1.000 & 1.000 & 1.000 & 1.000 \\ 
  * & 0.3 & 0.5 & 0.3 & * & * & 0.884 & 0.923 & 1.000 & 1.000 & 1.000 & 1.000 \\ 
  * & 0.1 & 0.5 & 0.3 & * & * & 0.148 & 0.158 & 0.286 & 0.290 & 0.546 & 0.560 \\ 
  0.1 & * & * & * & 0.5 & 0.3 & 0.180 & 0.184 & 0.301 & 0.298 & 0.559 & 0.601 \\ 
  0.3 & * & * & * & 0.5 & 0.3 & 0.939 & 0.933 & 1.000 & 1.000 & 1.000 & 1.000 \\ 
  0.5 & * & * & * & 0.5 & 0.3 & 1.000 & 1.000 & 1.000 & 1.000 & 1.000 & 1.000 \\ 
  * & 0.5 & * & * & 0.5 & 0.3 & 1.000 & 1.000 & 1.000 & 1.000 & 1.000 & 1.000 \\ 
  * & 0.3 & * & * & 0.5 & 0.3 & 0.879 & 0.917 & 0.998 & 0.999 & 1.000 & 1.000 \\ 
  * & 0.1 & * & * & 0.5 & 0.3 & 0.166 & 0.172 & 0.271 & 0.281 & 0.533 & 0.546 \\ 
\hline
\multicolumn{6}{l}{$\alpha=0.05$, \  $m_0=0$ $m_1=m_2=5$} & \multicolumn{2}{c}{$n=250$} & \multicolumn{2}{c}{$n=500$}& \multicolumn{2}{c}{$n=1000$} \\ \hline
  * & 0.5 & 0.5 & 0.4 & * & * & 0.210 & 0.166 & 0.191 & 0.190 & 0.210 & 0.255 \\ 
  * & 0.5 & * & * & 0.5 & 0.4 & 0.178 & 0.152 & 0.206 & 0.199 & 0.185 & 0.256 \\ 
  0.4 & * & 0.5 & 0.4 & * & * & 0.256 & 0.135 & 0.227 & 0.147 & 0.243 & 0.155 \\ 
  0.4 & * & * & * & 0.5 & 0.4 & 0.206 & 0.129 & 0.209 & 0.136 & 0.242 & 0.150 \\ 
  0.4 & 0.4 & 0.5 & 0.4 & * & * & 0.511 & 0.335 & 0.534 & 0.380 & 0.522 & 0.494 \\ 
  0.4 & 0.4 & * & * & 0.5 & 0.4 & 0.460 & 0.328 & 0.456 & 0.396 & 0.496 & 0.504 \\  \hline
	\end{tabular}
	\caption{Empirical power for generalized portmanteau tests in case of  mixed autocorrelation. DGPs are defined by the values of the parameters in the first six columns while the fitted models include only the seasonal components. In the upper block of the table $m_0=m_1=m_2=5$ are considered while in the lower block $m_0=0$ and $m_1=m_2=5$.} 
    \label{mixed_pow_47}
\end{small}
\end{table}

\section{\textcolor{black}{An application to road traffic data}}

\begin{figure}[H]
	\centering
	\includegraphics[width=0.48\linewidth]{"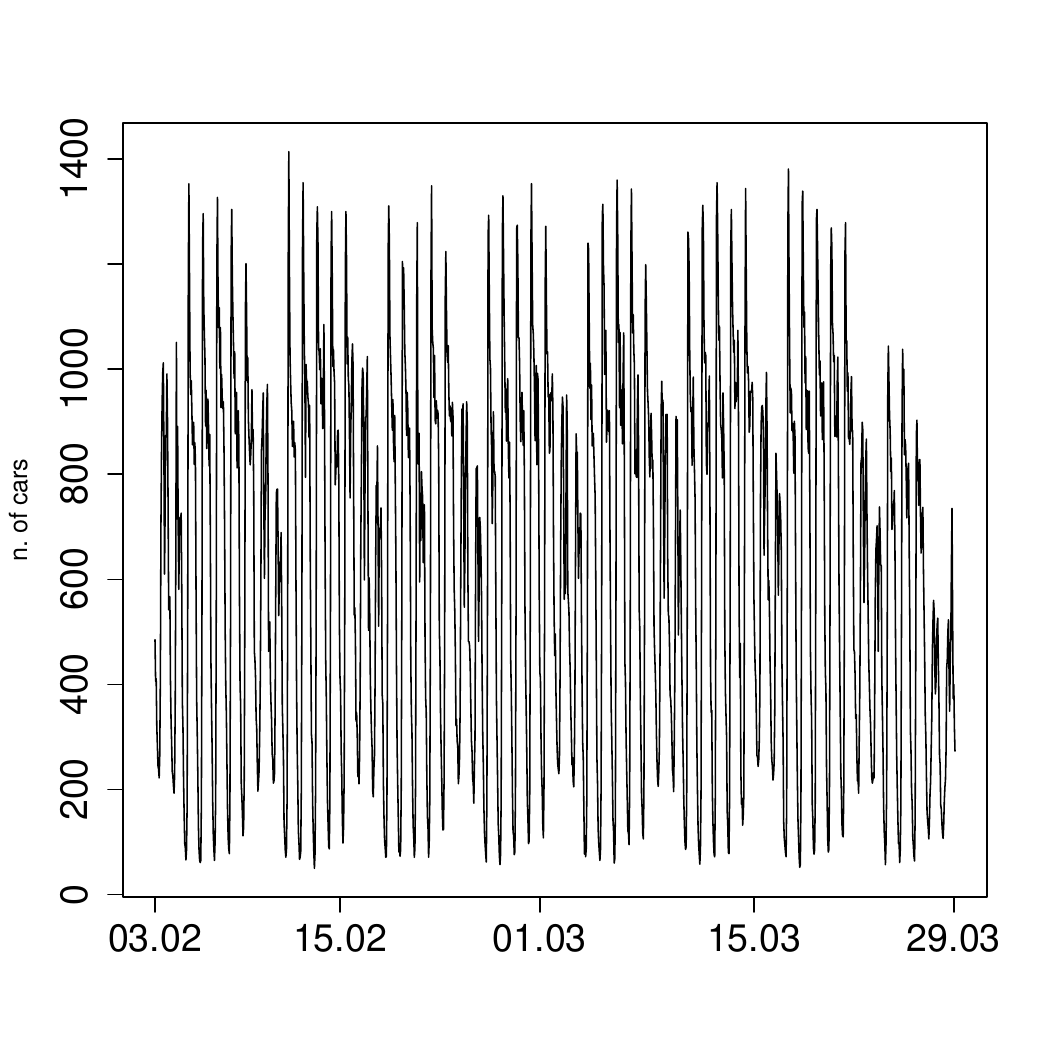"}	
	\includegraphics[width=0.48\linewidth]{"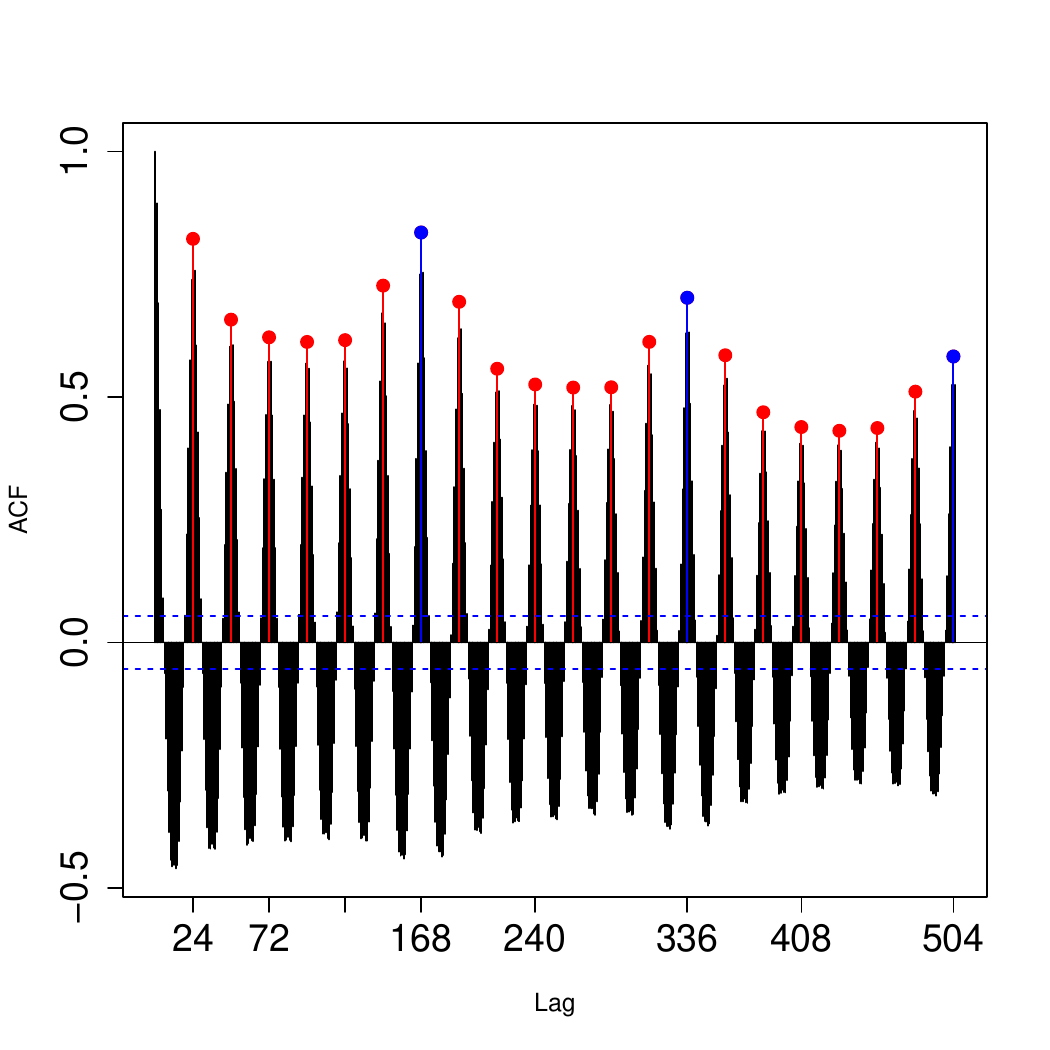"}
	\caption{Hourly time series of the road traffic on the Madrid M-40 ring road for the period from 3 February 2024 to 29 March 2024 (left) and its ACF (right). Red points indicate multiples of the daily period while blue points multiples of the weekly period.}
	\label{Madrid_1}
\end{figure}

To illustrate the usefulness of the proposed generalized portmanteau test statistics in an empirical setting, we consider an hourly time series of road traffic recorded at monitoring station No. 3573 in Madrid, which is part of a large-scale and extensive traffic detection system. This sensor, located on the M-40 ring road, was selected because it is one of the stations with the fewest missing observations.
\noindent Data are publicly available\footnote{Data can be downloaded from the following URL: 
{\tt https://datos.madrid.es/dataset/208627-0-}
{\tt transporte-ptomedida-historico/information}. }
at a 15-minute sampling frequency.\\
The time series considered spans eight weeks of hourly aggregated data, from 3 February 2024 to 29 March 2024, for a total of 1,344 observations. No missing values are present over this period.\\
Fig.~\ref{Madrid_1} displays the time series (left panel) and its sample autocorrelation function (right panel).
Since the data represent vehicular traffic flows in a large metropolitan area, it is natural to expect the presence of a daily seasonal component ($s_1 = 24$), reflecting the variation in traffic intensity across the hours of the day, a weekly seasonal component ($s_2 = 168$), associated with the different traffic patterns observed on the days of the week, and an annual seasonal component ($s_3 = 8760$), related to the time of year. However, because the data cover only a two-month period, the annual seasonal component can reasonably be neglected.\\
The daily and weekly seasonal patterns are clearly visible in both the time series plot and the sample autocorrelation function.

It was therefore decided to model the series using a multiplicative seasonal ARIMA (mSARIMA) model \citep{Lisi_Grigoletto}. Based on the statistical significance of the estimated parameters, the residual autocorrelation analysis and the Bayesian Information Criterion (BIC), the model that provided the best fit was an mSARIMA(1,0,2)(0,1,1)$_{24}$(0,1,1)$_{168}$ model, consisting of a double airline specification for seasonal components together with a short-term ARMA(1,2) component. The estimated model is reported in Table~\ref{Madrid_mod}.
\begin{table}[H]
\centering
\begin{tabular}{lcccc} \hline
	       & Estimate & Std. Error & z-value  &p-value \\ \hline    
	AR(1)   &  0.886 &   0.024  & 36.03  & $\approx 0$ \\
	MA(1)    &-0.375  &  0.040 & -9.27  & $\approx 0$ \\
	MA(2)     &-0.683  &  0.025 & -27.11  & $\approx 0$ \\
	SMA(1,1) &-0.091   & 0.038 & -2.41  &  0.016   \\
	SMA(2,1) &-0.561  &  0.026 & -20.81  & $\approx 0$ \\	\hline
\end{tabular}
\caption{Estimated model parameters and their $p$-values.
$\hat \sigma_\varepsilon^2 = 4006.27$  log-lik = -5365.39.}
\label{Madrid_mod}
\end{table}

\begin{figure}[H]
	\centering
	\includegraphics[width=0.49\linewidth]{"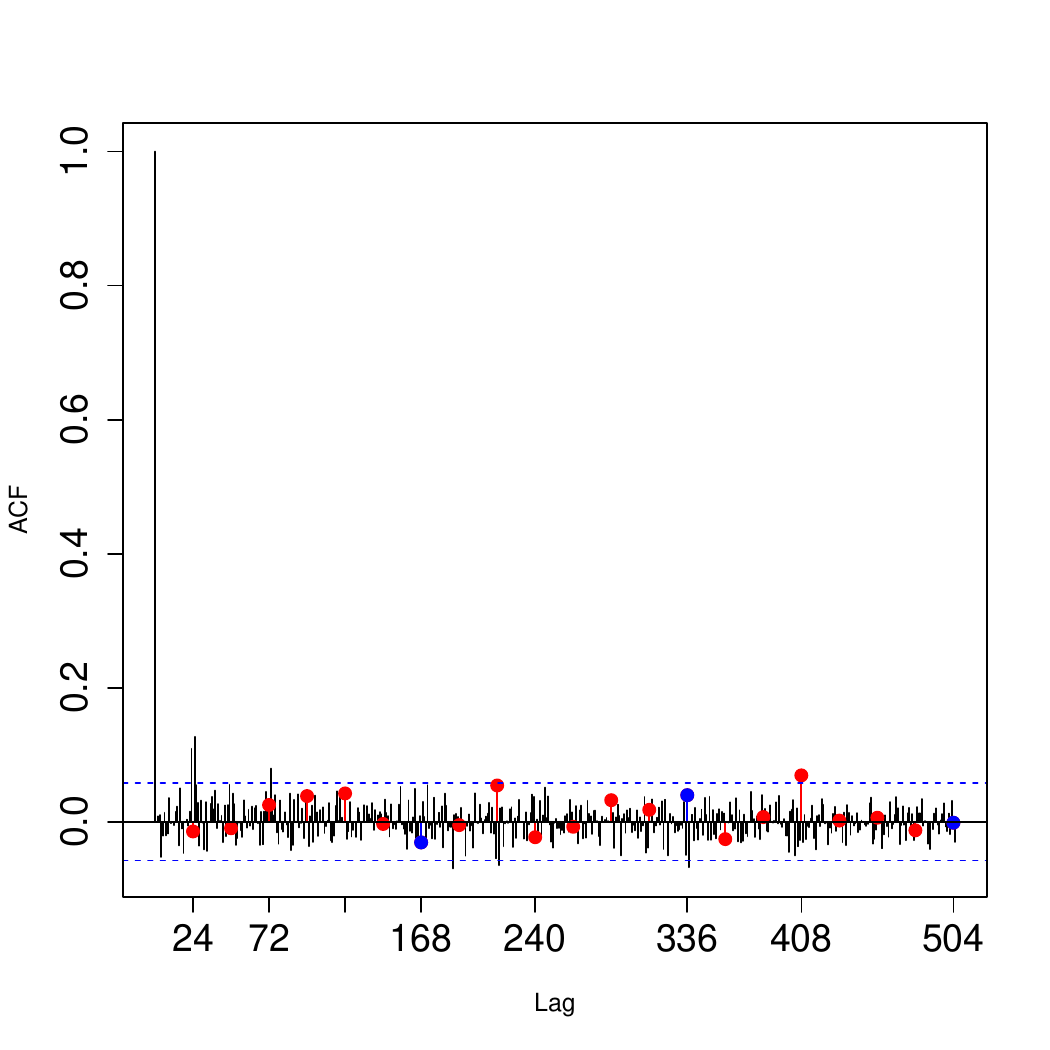"}	
	\includegraphics[width=0.49\linewidth]{"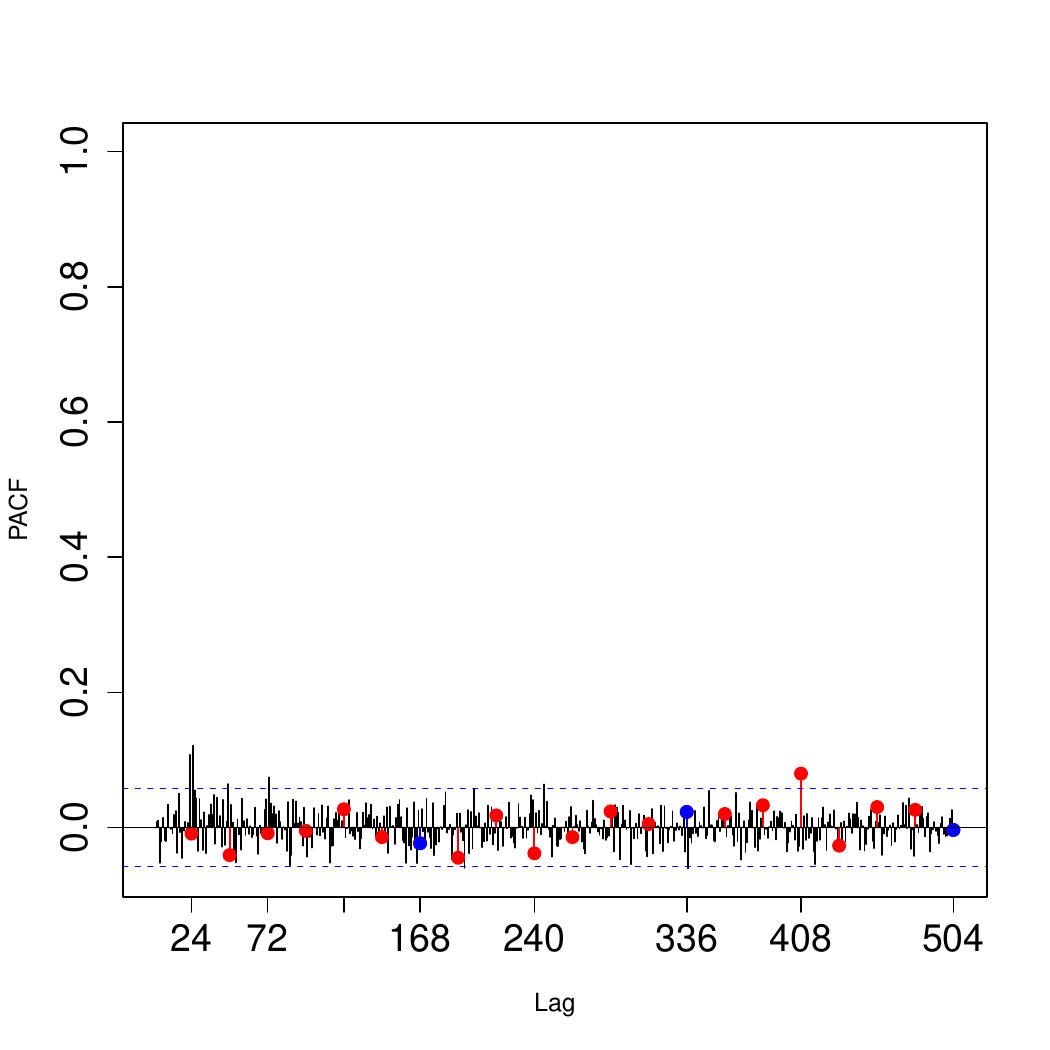"}
	\caption{ACF (left) and PACF (right) of the estimated model. Red points indicate multiples of the daily period while blue points multiples of the weekly period.}
	\label{Madrid_2}
\end{figure}

The autocorrelation functions of the model's residuals are given in Fig.~\ref{Madrid_2}. It is clear that most of the linear dependence on data has been explained, but since some lags are still outside of the Bartlett's bands, some tests for no residual autocorrelation are suitable. Table \ref{Madrid_test} lists the results of the application of the generalized Ljung-Box and Monti tests for $m_0=8$, $m_1=5$ and $m_2=5$. Neither test rejects\footnote{\textcolor{black}{The choice of the specific values of $m_0$, $m_1$ and $m_2$ is not critical and the same conclusions would have been reached considering slightly different values for the lags.}} the hypothesis of no residual short-term or periodic autocorrelation both for the daily cycle and for the weekly cycle, confirming the adequacy of the estimated model.

\begin{table}[H]
	\centering
	\begin{tabular}{lccc} \hline
		Test & Statistics & df  & $p$-value \\ \hline    
		G-LB   &  16.42 &  13  & 0.227 \\
		G-M   &  11.90 &  13  & 0.535 \\	\hline
	\end{tabular}
	\caption{Generalized portmanteau test on the model residuals for $m_0=8$, $m_1=5$ and $m_2=5$}
	\label{Madrid_test}
\end{table}

\section{Concluding remarks}

This article proposes a generalized version of the classical and seasonal portmanteau test for univariate time series. The extension allows one to correctly address two issues not yet sufficiently covered in the literature: the case of multiple seasonal residual autocorrelation and that of mixed (short-term and seasonal) residual autocorrelation.\\
In particular, we focused on the generalized version of the \cite{ljungbox} and~\cite{Monti_1994} tests.
We did not analyze the performance of the generalization for all tests present in the
literature, as e.g.\ the \cite{fishergallagher},
\cite{pena_rodriguez2006} and \cite{mahdimcleod} tests. However, the generalized approach suggested in this work can also be easily implemented in those cases \textcolor{black}{and can be a topic for future research.}\\
The test statistics and their asymptotic distributions were straightforwardly derived from the original tests and their empirical properties were explored.\\
We studied, by simulation, the type-I error and the power of the generalized portmanteau tests for multi-seasonal time series and to detect mixed residual autocorrelation. 
In addition, the asymptotic distributions of the tests were verified using statistical tests. \\
In all the considered cases, the results are clear and coherent: 
the proposed generalized approach works well and can be easily
extended to most of the existing tests.
In general, we recommend the use of the generalized Ljung-Box and
Monti tests, which are simple to apply and are not particularly affected by the number of lags or by the sample size. For these reasons, they can be considered reference tests to check the goodness-of-fit within the class of mSARIMA models \citep{Lisi_Grigoletto}.\\
Even if in this paper we always considered time series with no more than two seasonal components, which is often the case of interest, the proposed extension is not limited to a double seasonality and, in principle, works for any number of cyclical components. Clearly, in practice, an increasing number of periodicities also increases the occurrence of possible computational issues. \\
\textcolor{black}{We plan to release an R package on CRAN providing functions for fitting mSARIMA models and performing generalized portmanteau tests. Until then, these functions are available from the authors upon request.}

\bibliographystyle{apalike}
\bibliography{acf_biblio}

\section*{\textcolor{black}{Appendix A}}

We follow the derivation strategy of \cite{McLeod_1978}, which proceeds through three
lemmas culminating in a martingale central limit argument. The only genuinely new
step needed to cover the multi-seasonal mSARMA models of~\cite{Lisi_Grigoletto}
is the computation of the score derivatives $\partial \hat \varepsilon_t/\partial \beta_i$
when the autoregressive and moving-average operators are \emph{products} of several
seasonal factors rather than a single polynomial. We isolate this step as Lemma~A
below; the rest of the argument is unchanged from \cite{McLeod_1978}.

Consider the mSARMA model by~\cite{Lisi_Grigoletto} defined in equation~(\ref{msarima}). For
simplicity we take the model to be already stationary, i.e.\
$d=D_i=0$; the integrated case is handled as usual by first
differencing. Write
\[
\widetilde\Phi(B)=\phi(B)\prod_{i=1}^{n_s}\Phi_i(B^{S_i}), \qquad
\widetilde\Theta(B)=\theta(B)\prod_{i=1}^{n_s}\Theta_i(B^{S_i})
\]
for the overall AR and MA operators, and let
\[
\bm\beta=\big(\phi_1,\dots,\phi_p,\ \theta_1,\dots,\theta_q,\
\Phi_{1,1},\dots,\Phi_{1,P_1},\dots,\Phi_{n_s,1},\dots,\Phi_{n_s,P_{n_s}},\
\Theta_{1,1},\dots,\Theta_{n_s,Q_{n_s}}\big)
\]
be the $\eta$-dimensional parameter vector, with
\[
\eta = p+q+\sum_{i=1}^{n_s}(P_i+Q_i).
\]

Given least-squares (or Gaussian ML) estimates $\hat{\bm\beta}$, let $\hat \varepsilon_t$ be the
corresponding estimated white-noise series, and for a fixed vector of lags
${\bf L}=(l_1,\dots,l_k)$ (which may mix non-seasonal lags and lags that are multiples of
one or more periods $S_i$) define the residual autocorrelation vector
$\widehat{\rho}({\bf L})=\big(\hat \rho_{l_1},\dots,\hat \rho_{l_k}\big)^T$, with
$\hat \rho_l$ defined in equation~(\ref{eq:rhohat}).

\paragraph{Lemma A.}{
\emph{Define the auxiliary processes}
\[
\phi(B)\,v_t=-\varepsilon_t, \qquad
\theta(B)\,u_t=\varepsilon_t, \qquad
\Phi_i(B^{S_i})\,V_{i,t}=-\varepsilon_t, \qquad
\Theta_i(B^{S_i})\,U_{i,t}=\varepsilon_t
\qquad (i=1,\dots,n_s)\ .
\]
\emph{Then, apart from terms} $O_p(1/\sqrt n)$,
\[
\frac{\partial \hat \varepsilon_t}{\partial \phi_j}=v_{t-j}\ \ (1\le j\le p),
\qquad
\frac{\partial \hat \varepsilon_t}{\partial \theta_j}=u_{t-j}\ \ (1\le j\le q),
\]
\[
\frac{\partial \hat \varepsilon_t}{\partial \Phi_{i,k}}=V_{i,\,t-kS_i}\ \ (1\le k\le P_i),
\qquad
\frac{\partial \hat \varepsilon_t}{\partial \Theta_{i,k}}=U_{i,\,t-kS_i}\ \ (1\le k\le Q_i).
\]
}

\begin{proof}
Consider the equation
\[
\widetilde\Theta(B;\bm\beta)\,\hat \varepsilon_t(\bm\beta) = \widetilde\Phi(B;\bm\beta)\, Y_t\ .
\]

Differentiating both sides\footnote{Note that $\hat\varepsilon_t(\bm\beta)$ is a filtered version of $Y_t$ for any trial $\bm\beta$; the subsequent substitution $Y_t = \widetilde\Phi(B)^{-1} \widetilde\Theta(B)\varepsilon_t$ uses the \emph{true} operators, so implicitly the derivative is evaluated at $\bm\beta=\bm\beta_{\text{true}}$, consistently with the $O_p(1/\sqrt n)$ in the Lemma's statement.} with respect to a generic component $\beta_g$ of
$\bm\beta$ (product rule on the operators; $Y_t$ does not depend on $\bm\beta$):
\[
\frac{\partial \widetilde\Theta(B)}{\partial \beta_g}\,\hat \varepsilon_t
+ \widetilde\Theta(B)\,\frac{\partial \hat \varepsilon_t}{\partial \beta_g}
= \frac{\partial \widetilde\Phi(B)}{\partial \beta_g}\, Y_t\ .
\]
Consider first the case $\beta_g=\Phi_{i,k}$. Then, since
$\partial\widetilde\Theta/\partial\Phi_{i,k}=0$,
\begin{equation}
  \label{eq:partder}
  \widetilde\Phi(B)\ \frac{\partial \hat \varepsilon_t}{\partial \Phi_{i,k}}
  = \frac{\partial \widetilde\Phi(B)}{\partial \Phi_{i,k}}\, \varepsilon_t\ ,
  \tag{$A.1$}
\end{equation}
where we substituted $Y_t$ with $\widetilde\Phi(B)^{-1}\widetilde\Theta(B)\,\varepsilon_t$.

Now, since $\widetilde\Phi(B)=\phi(B)\prod_i\Phi_i(B^{S_i})$,
\[
\frac{\partial \widetilde\Phi(B)}{\partial \Phi_{i,k}}
= -B^{kS_i}\,\phi(B)\prod_{i'\neq i}\Phi_{i'}(B^{S_{i'}})\ .
\]
Dividing both sides of equation~($A.1$) by $\widetilde\Phi(B)=\phi(B)\,\Phi_i(B^{S_i})\prod_{i'\neq i}\Phi_{i'}(B^{S_{i'}})$,
every factor except $\Phi_i(B^{S_i})$ cancels, leaving
\[
\frac{\partial \hat \varepsilon_t}{\partial \Phi_{i,k}}
= -\frac{B^{kS_i}}{\Phi_i(B^{S_i})}\,\varepsilon_t = V_{i,\,t-kS_i}\ .
\]
The derivations for $\phi_j$, $\theta_j$ and $\Theta_{i,k}$ are identical, mutatis
mutandis, using $\widetilde\Theta(B)=\theta(B)\prod_i\Theta_i(B^{S_i})$ for the MA side.
\end{proof}

\noindent
The essential point is that the derivative with
respect to any seasonal-block parameter depends \emph{only on its own factor}
$\Phi_i(B^{S_i})$ or $\Theta_i(B^{S_i})$: the other $n_s-1$ seasonal factors and the
non-seasonal part cancel out of the ratio, exactly as in McLeod's (1978) original
single-factor computation (his equations (26)--(27)).

\paragraph{Information matrix and the $\mathbf X_L$ matrix.}{
Let $\Fisher$ be the $\eta\times\eta$ information matrix, partitioned according to the blocks
$\{\phi,\theta,\Phi_1,\dots,\Phi_{n_s},\Theta_1,\dots,\Theta_{n_s}\}$, with $(a,b)$
entries given by the auto- and cross-covariances
\[
\gamma_{\xi\zeta}(k)=E(\xi_t\,\zeta_{t+k}),
\qquad \xi,\zeta\in\{v,u,V_1,\dots,V_{n_s},U_1,\dots,U_{n_s}\},
\]
generalizing McLeod's equations (9)--(12) and (41).

For each lag $l\in L$ define the impulse functions coefficients
\[
\frac{1}{\phi(B)}=\sum_{r=0}^\infty \phi'_rB^r, \qquad
\frac{1}{\theta(B)}=\sum_{r=0}^\infty \theta'_rB^r, \qquad
\frac{1}{\Phi_i(B^{S_i})}=\sum_{r=0}^\infty \Phi'_{i,r}B^{rS_i}, \qquad
\frac{1}{\Theta_i(B^{S_i})}=\sum_{r=0}^\infty \Theta'_{i,r}B^{rS_i},
\]
and let $\mathbf X_L$ be the $k\times\eta$ matrix whose row corresponding to lag $l$ is
\[
\Big(-\phi'_{l-j}\ \Big|\ \theta'_{l-j}\ \Big|\
-\Phi'_{1,\,l-kS_1}\ \Big|\cdots\Big|\ -\Phi'_{n_s,\,l-kS_{n_s}}\ \Big|\
\Theta'_{1,\,l-kS_1}\ \Big|\cdots\Big|\ \Theta'_{n_s,\,l-kS_{n_s}}\Big).
  \  \  \tag{$A.2$}
   \]
}

\paragraph{Proof of Theorem 1.}{
The proof follows McLeod's (1978) three-lemma scheme verbatim, with Lemma~A above
replacing his single-operator score computation. Define the white
noise autocorrelations as
\[
  r_l=\sum_{t=1}^{n-l} \varepsilon_{t}\ \varepsilon_{t-l}/
  \sum_{t=1}^{n} \varepsilon_{t}^2 \hspace{0.7cm}(l=1,2,\ldots)\ .
\]

\emph{Step 1 (analogue of McLeod's Lemma 1).} For any fixed $l$,
$r_l=c_l+O_p(1/n)$ with $c_l=\sum_t \varepsilon_t\,\varepsilon_{t+l}/n$; this follows from a Taylor
expansion of $r_l$ as a function of $(c_l,c_0)$ about $(0,1)$ and is unaffected by
the multiplicative seasonal structure of the model.

\emph{Step 2 (analogue of McLeod's Lemma 2).} Let $\hat{\bm\beta}$ minimize
$S(\bm\beta)=\sum_t \hat \varepsilon_t^2$. A Taylor expansion of $\partial S/\partial\bm\beta$
about $\bm\beta$ evaluated at $\hat{\bm\beta}$, together with
$\tfrac1n\,\partial^2S/\partial\bm\beta\partial\bm\beta^T=2\,\Fisher+O_p(1/\sqrt n)$,
gives
\[
\hat{\bm\beta}-\bm\beta = \Fisher^{-1}\mathbf s_c + O_p(1/n),
\]
where the $i$th component of $\mathbf s_c$ is $-\sum_t \varepsilon_t\xi_{t-i}/n$ for the
appropriate auxiliary process $\xi\in\{v,u,V_i,U_i\}$ from Lemma~A, according to which
block of $\bm\beta$ the index $i$ belongs to.

\emph{Step 3 (analogue of McLeod's Lemma 3).} Define ${r(\bf L)} =
(r_{l_1},\ldots,r_{l_k})^T$. By Steps 1--2, every linear combination
of the entries of $(\hat{\bm\beta}-\bm\beta,\ r(\bf L))$ is, apart from terms
$O_p(1/n)$, an average of a sequence of martingale differences in $\varepsilon_t$; hence, by the
martingale central limit theorem \citep{Billingsley_1961}, $\sqrt n$ times this linear
combination is asymptotically normal, so that $\sqrt n(\hat{\bm\beta}-\bm\beta,\
r(\bf L))$ is asymptotically jointly normal. The relevant cross-covariances are, by
the same fourth-moment argument used by McLeod (\citealp{Hannan_1970}, p.\ 23),
\[
\lim_{n\to\infty} n\,E\big(s_{c,\Phi_{i,k}}\,c_l\big) = -\Phi'_{i,\,l-kS_i}
\]
for the AR seasonal parameters of factor $i$, with the analogous relations
$\lim_n n\,E(s_{c,\phi_j}c_l)=-\phi'_{l-j}$, $\lim_n n\,E(s_{c,\theta_j}c_l)=\theta'_{l-j}$
and $\lim_n n\,E(s_{c,\Theta_{i,k}}c_l)=\Theta'_{i,\,l-kS_i}$ holding for the
non-seasonal and seasonal MA blocks. Collecting all blocks, the asymptotic covariance
matrix of $\sqrt n\, (\hat{\bm\beta}-\bm\beta)$ and $\sqrt n\, r(\bf L)$ is
$-\Fisher^{-1}\mathbf X_L^T$.

\emph{Conclusion.} A further Taylor expansion of $r(\bf L)$ about $\bm\beta=\hat{\bm\beta}$
gives, as in McLeod's proof of Theorem 1,
\[
\widehat{\rho}({\bf L}) = r({\bf L}) + X_L\,(\hat{\bm\beta}-\bm\beta) + O_p(1/n)\ .
\]
Combining this with Step 3 yields the stated asymptotic covariance matrix of
$\widehat{\rho}(\bf L)$.
}

\noindent
Theorem~1 contains as special cases:
\begin{itemize}
\item McLeod's (1978) Theorem 1, when $n_s=0$;
\item McLeod's (1978) Theorem 2 (single SARMA seasonality), when $n_s=1$;
\item the purely multi-seasonal case, obtained by
setting $p=q=0$, $\eta=\sum_i(P_i+Q_i)$, and
$L=\{S_1,\dots,m_1S_1\}\cup\cdots\cup\{S_{n_s},\dots,m_{n_s}S_{n_s}\}$;
\item the mixed short-term/seasonal case, with
$L=\{1,\dots,m_0\}\cup\{S_1,\dots,m_1S_1\}\cup\cdots\cup\{S_{n_s},\dots,m_{n_s}S_{n_s}\}$.
\end{itemize}

\section*{\textcolor{black}{Appendix B}}

\begin{table}[H]
	\centering
    \begin{small}
	\begin{tabular}{cc|cc|cc|cc}
		\hline
$\Phi_{1,1}$ & $\Phi_{2,1}$ & G-LB & G-M  & G-LB & G-M  & G-LB & G-M  \\ \hline
      \multicolumn{2}{l}{$\alpha=0.01$}    & \multicolumn{2}{c}{$n=250$}  & \multicolumn{2}{c}{$n=500$} & \multicolumn{2}{c}{$n=1000$}  \\ \hline 
  0.200 & 0.200 & 0.013 & 0.009 & 0.011 & 0.014 & 0.011 & 0.010 \\ 
  0.500 & 0.200 & 0.011 & 0.006 & 0.011 & 0.009 & 0.010 & 0.009 \\ 
  0.800 & 0.200 & 0.009 & 0.009 & 0.011 & 0.009 & 0.006 & 0.006 \\ 
  0.200 & 0.500 & 0.006 & 0.011 & 0.011 & 0.011 & 0.007 & 0.007 \\ 
  0.500 & 0.500 & 0.012 & 0.009 & 0.010 & 0.012 & 0.012 & 0.011 \\ 
  0.800 & 0.500 & 0.005$^{*}$ & 0.004$^{**}$ & 0.011 & 0.007 & 0.009 & 0.007 \\ 
  0.200 & 0.800 & 0.007 & 0.010 & 0.010 & 0.010 & 0.006 & 0.007 \\ 
  0.500 & 0.800 & 0.013 & 0.012 & 0.008 & 0.007 & 0.014 & 0.013 \\ 
  0.800 & 0.800 & 0.011 & 0.007 & 0.017$^{**}$ & 0.011 & 0.014 & 0.014 \\ 
 \hline
      \multicolumn{2}{l}{$\alpha=0.05$}    & \multicolumn{2}{c}{$n=250$}  & \multicolumn{2}{c}{$n=500$} & \multicolumn{2}{c}{$n=1000$}  \\ \hline 
   0.200 & 0.200 & 0.044 & 0.051 & 0.050 & 0.050 & 0.045 & 0.052 \\ 
  0.500 & 0.200 & 0.046 & 0.049 & 0.049 & 0.052 & 0.046 & 0.051 \\ 
  0.800 & 0.200 & 0.048 & 0.042 & 0.048 & 0.048 & 0.038$^{*}$ & 0.039$^{*}$ \\ 
  0.200 & 0.500 & 0.052 & 0.046 & 0.043 & 0.045 & 0.048 & 0.051 \\ 
  0.500 & 0.500 & 0.050 & 0.048 & 0.048 & 0.052 & 0.050 & 0.051 \\ 
  0.800 & 0.500 & 0.038$^{**}$ & 0.042 & 0.050 & 0.048 & 0.048 & 0.050 \\ 
  0.200 & 0.800 & 0.042 & 0.044 & 0.050 & 0.051 & 0.044 & 0.041 \\ 
  0.500 & 0.800 & 0.054 & 0.052 & 0.045 & 0.043 & 0.052 & 0.056 \\ 
  0.800 & 0.800 & 0.059 & 0.046 & 0.065$^{**}$ & 0.063$^{**}$ & 0.070$^{**}$ & 0.069$^{**}$ \\ 
    \hline
      \multicolumn{2}{l}{$\alpha=0.10$}    & \multicolumn{2}{c}{$n=250$}  & \multicolumn{2}{c}{$n=500$} & \multicolumn{2}{c}{$n=1000$}  \\ \hline 
  0.200 & 0.200 & 0.098 & 0.098 & 0.095 & 0.102 & 0.096 & 0.104 \\ 
  0.500 & 0.200 & 0.100 & 0.108 & 0.104 & 0.104 & 0.103 & 0.100 \\ 
  0.800 & 0.200 & 0.093 & 0.102 & 0.082 & 0.096 & 0.088 & 0.088 \\ 
  0.200 & 0.500 & 0.106 & 0.098 & 0.087 & 0.090 & 0.106 & 0.102 \\ 
  0.500 & 0.500 & 0.092 & 0.102 & 0.104 & 0.106 & 0.090 & 0.090 \\ 
  0.800 & 0.500 & 0.082$^{**}$ & 0.078$^{**}$ & 0.102 & 0.094 & 0.098 & 0.106 \\ 
  0.200 & 0.800 & 0.087 & 0.089 & 0.096 & 0.107 & 0.088 & 0.087 \\ 
  0.500 & 0.800 & 0.104 & 0.106 & 0.088 & 0.089 & 0.106 & 0.107 \\ 
  0.800 & 0.800 & 0.116 & 0.090 & 0.138$^{**}$ & 0.120$^{**}$ & 0.128$^{**}$ & 0.124$^{**}$ \\ 
   \hline
	\end{tabular}
    	\caption{Multi-seasonal case: empirical level ($\alpha_{obs}$) of the test. Data generated from an mSARIMA$(0,0,0)(1,0,0)_4 (1,0,0)_{12}$ model for different values of parameters $\Phi_{1,1}$ and $\Phi_{2,1}$ and sample sizes. G-LB=generalized Ljung-Box test, G-M=generalized Monti test, $\alpha$=nominal test level, $N=2000$, $\bL=(4, 8, 12, 16, 20, 24, 36, 48, 60)$, $(S_1,S_2)=(4,12)$. \\ $^*$  and $^{**}$ denote empirical levels significantly different from the nominal one at $5\%$ and at $1\%$, respectively.}  \label{multis_level_sarsar_412}
        \end{small}
\end{table}

\begin{table}[H]
	\centering
        \begin{small}
	\begin{tabular}{cc|cc|cc|cc} \hline
$\Theta_{1,1}$ & $\Theta_{2,1}$ & G-LB & G-M  & G-LB & G-M & G-LB & G-M  \\ \hline
      \multicolumn{2}{l}{$\alpha=0.01$}    & \multicolumn{2}{c}{$n=250$}  & \multicolumn{2}{c}{$n=500$} & \multicolumn{2}{c}{$n=1000$}  \\ \hline 
 0.200 & 0.200 & 0.014 & 0.009 & 0.013 & 0.013 & 0.011 & 0.007 \\ 
  0.500 & 0.200 & 0.012 & 0.012 & 0.011 & 0.012 & 0.011 & 0.013 \\ 
  0.800 & 0.200 & 0.009 & 0.007 & 0.009 & 0.006 & 0.007 & 0.010 \\ 
  0.200 & 0.500 & 0.010 & 0.009 & 0.010 & 0.013 & 0.011 & 0.012 \\ 
  0.500 & 0.500 & 0.008 & 0.010 & 0.009 & 0.010 & 0.014 & 0.009 \\ 
  0.800 & 0.500 & 0.016$^{*}$ & 0.010 & 0.011 & 0.014 & 0.012 & 0.009 \\ 
  0.200 & 0.800 & 0.010 & 0.013 & 0.014 & 0.013 & 0.012 & 0.014 \\ 
  0.500 & 0.800 & 0.016$^{*}$ & 0.016$^{*}$ & 0.014 & 0.013 & 0.014 & 0.012 \\ 
  0.800 & 0.800 & 0.020$^{**}$ & 0.012 & 0.025$^{**}$ & 0.025$^{**}$ & 0.028$^{**}$ & 0.021$^{**}$ \\ 
 \hline
      \multicolumn{2}{l}{$\alpha=0.05$}    & \multicolumn{2}{c}{$n=250$}  & \multicolumn{2}{c}{$n=500$} & \multicolumn{2}{c}{$n=1000$}  \\  
 \hline
0.200 & 0.200 & 0.052 & 0.054 & 0.053 & 0.058 & 0.051 & 0.057 \\ 
 0.500 & 0.200 & 0.047 & 0.054 & 0.048 & 0.049 & 0.059 & 0.057 \\ 
  0.800 & 0.200 & 0.042 & 0.045 & 0.040 & 0.042 & 0.049 & 0.055 \\ 
  0.200 & 0.500 & 0.051 & 0.051 & 0.050 & 0.054 & 0.051 & 0.052 \\ 
  0.500 & 0.500 & 0.045 & 0.044 & 0.048 & 0.054 & 0.044 & 0.050 \\ 
  0.800 & 0.500 & 0.052 & 0.055 & 0.054 & 0.052 & 0.047 & 0.047 \\ 
  0.200 & 0.800 & 0.048 & 0.048 & 0.058 & 0.055 & 0.050 & 0.052 \\ 
  0.500 & 0.800 & 0.065$^{**}$ & 0.066$^{**}$ & 0.057 & 0.054 & 0.059 & 0.056 \\ 
  0.800 & 0.800 & 0.080$^{**}$ & 0.078$^{**}$ & 0.086$^{**}$ & 0.087$^{**}$ & 0.084$^{**}$ & 0.082$^{**}$ \\ 
 \hline
      \multicolumn{2}{l}{$\alpha=0.10$}    & \multicolumn{2}{c}{$n=250$}  & \multicolumn{2}{c}{$n=500$} & \multicolumn{2}{c}{$n=1000$}  \\ \hline 
 0.200 & 0.200 & 0.101 & 0.112 & 0.098 & 0.113 & 0.112 & 0.112 \\ 
  0.500 & 0.200 & 0.100 & 0.102 & 0.092 & 0.103 & 0.113 & 0.106 \\ 
  0.800 & 0.200 & 0.087 & 0.095 & 0.088 & 0.086$^{*}$ & 0.102 & 0.110 \\ 
  0.200 & 0.500 & 0.100 & 0.105 & 0.100 & 0.104 & 0.106 & 0.106 \\ 
  0.500 & 0.500 & 0.092 & 0.100 & 0.091 & 0.100 & 0.097 & 0.102 \\ 
  0.800 & 0.500 & 0.102 & 0.105 & 0.091 & 0.105 & 0.098 & 0.098 \\ 
  0.200 & 0.800 & 0.102 & 0.104 & 0.110 & 0.111 & 0.102 & 0.098 \\ 
  0.500 & 0.800 & 0.124$^{**}$ & 0.124$^{**}$ & 0.106 & 0.104 & 0.113 & 0.118$^{**}$ \\ 
  0.800 & 0.800 & 0.192$^{**}$ & 0.162$^{**}$ & 0.180$^{**}$ & 0.155$^{**}$ & 0.199$^{**}$ & 0.188$^{**}$ \\  
   \hline
	\end{tabular}
	\caption{Multi-seasonal case: empirical level ($\alpha_{obs}$)  of the test. Data generated from mSARIMA$(0,0,0)(0,0,1)_4 (0,0,1)_{12}$ model for different values of parameters $\Theta_{1,1}$ and $\Theta_{2,1}$ and sample sizes. G-LB=generalized Ljung-Box test, G-M=generalized Monti test, $\alpha$=nominal significance level, $N=2000$, $\bL=(4, 8, 12, 16, 20, 24, 36, 48, 60)$, $(S_1,S_2)=(4,12)$.
	$^*$  and $^{**}$ denote empirical levels significantly different from the nominal one at $5\%$ and at $1\%$, respectively.} \label{multis_level_smasma_412}
    \end{small}
\end{table}

\end{document}